\documentclass[fleqn,11pt]{wlscirep}

\usepackage[utf8]{inputenc}
\usepackage[T1]{fontenc}
\usepackage{hyperref}
\usepackage{authblk}
\usepackage{xcolor}
\definecolor{hgrevision}{RGB}{112,48,160}

\newsavebox{\termoldbox}

\usepackage{siunitx}
\usepackage{ragged2e}
\usepackage{algorithm}
\usepackage{algpseudocode}
\usepackage{amsthm}
\usepackage{tikz}
\usepackage{cleveref}
\crefname{figure}{Fig.}{Figs.}
\usetikzlibrary{arrows.meta,calc,positioning,decorations.pathreplacing}
\usepackage{mathtools}

\usepackage[version=4]{mhchem}

\usepackage{booktabs}
\usepackage{array}
\usepackage{amsmath,amssymb,bm}
\allowdisplaybreaks

\newcommand{\dd}{\mathrm{d}}

\newcommand{\E}{\mathbb{E}}

\usepackage{graphicx}
\usepackage{subcaption}
\newcommand{\name}{Pharos}

\usepackage{caption}

\providecommand{\authorcheck}[1]{{\color{orange}\par\noindent\textbf{Further check.} \textit{#1}\par}}

\newcommand{\Prob}{\mathbb P}
\newcommand{\Lop}{\mathcal L}

\usepackage{comment}
\title{\name: Learning to Control Rare Events at Scale}

\author[1,2,*,$\#$]{Jiajun He}
\author[3,4,*]{Jacky Chen}
\author[5,6,*]{Weiliang Luo}
\author[5,6]{Heather Kulik}
\author[4]{David R. Koes}
\author[1,$\dagger$]{Carles Domingo-Enrich}
\author[1,$\dagger$]{Yuanqi Du}

\affil[1]{Microsoft Research New England, Cambridge, MA 02142, USA}
\affil[2]{Department of Engineering, University of Cambridge, Cambridge, CB2 1PZ, UK}
\affil[3]{CMU-Pitt Joint Program in Computational Biology, Pittsburgh, PA 15213, USA}
\affil[4]{Department of Computational and Systems Biology, University of Pittsburgh, Pittsburgh, PA 15213, USA}
\affil[5]{Department of Chemistry, Massachusetts Institute of Technology, Cambridge, MA 02139, USA}
\affil[6]{Department of Chemical Engineering, Massachusetts Institute of Technology, Cambridge, MA 02139, USA}

\affil[*]{These authors contributed equally.}
\affil[$\#$]{Work done during internship at Microsoft Research.}
\affil[$\dagger$]{Correspondence. Email: carlesd@microsoft.com, yuanqidu@microsoft.com}

\begin{document}

\begin{abstract}
Rare events drive transformations across physics, chemistry and biology, yet predicting their mechanisms and rates remains a fundamental challenge. Here we introduce \name{}, a framework that learns to make rare transitions accessible while predicting how frequently they occur under physical dynamics. Through stochastic optimal control, Pharos jointly learns the probability of completing a transition and how to steer simulations toward it, allowing sampling and learning to improve one another. Across biomolecular benchmarks, Pharos learns accurate transition probabilities with order-of-magnitude gains in convergence speed over previous algorithms. It transfers across ion assembly systems in explicit solvent while maintaining accurate reaction rate predictions, scales to host-guest binding systems that span millisecond timescales, and enables chemical reaction studies with a computationally demanding machine learning interatomic potential. By combining efficient sampling, quantitative kinetics, and transferability, Pharos opens a route toward foundation models of rare event mechanisms and rates across molecular systems.
\end{abstract}

\maketitle

\section{Introduction}

Rare events underlie many important transformations in nature, including protein conformational changes, chemical reactions, and phase transitions\cite{geissler2001autoionization,wolde1997enhancement,noe2009constructing,lupi2017role}. Their rates determine whether these transformations occur on experimentally accessible timescales and how competing processes shape their outcomes. Predicting these rates remains a central challenge because systems can spend exceedingly long periods fluctuating within metastable basins before undergoing a transition\cite{lindorff2011fast}. Direct simulations therefore devote most of their computational effort to waiting, with few observations of the events that determine the kinetics. Enhanced sampling has greatly expanded access to transition pathways, but observing how a transformation occurs does not by itself establish how frequently it occurs under the physical dynamics. Consequently, advances in mechanistic understanding are not always accompanied by quantitative predictions of reaction rates.

A central link between microscopic mechanisms and macroscopic kinetics is the committor function: the probability that a system starting from a given state reaches the product basin before the reactant basin\cite{onsager1938initial}. Widely regarded as the ideal reaction coordinate\cite{du1998transition,bolhuis2000reaction,best2005reaction}, the committor measures progress toward a transition and identifies configurations equally likely to reach either basin. Within transition path theory, it also enters expressions for reactive probability currents and reaction rates, providing a principled route from molecular configurations to quantitative kinetics \cite{vanden2006towards}. This connection has motivated substantial efforts to learn the committor using molecular simulations and machine learning, including direct trajectory shooting, variational principles, and dynamical self-consistency relations \cite{weare2026machine,hummer2004transition,jung2023machine,li2020solving,li2019computing,chen2023discovering,li2022semigroup,strahan2021long,mitchell2024committor}.

These approaches share a fundamental difficulty: obtaining informative observations of the reactive region. Unbiased simulations predominantly visit metastable basins, where the outcome is largely determined, while rarely exploring the intervening configurations where commitment to either basin changes. Sampling and committor learning are therefore closely interdependent. An accurate committor can guide simulations toward informative configurations, while those simulations provide the evidence needed to improve its estimate \cite{rotskoff2022active,trizio2025everything,kang2024computing}. A common strategy is to embed transition path sampling\cite{dellago2002transition} within an iterative learning loop, using sampled trajectories to refine the committor and the learned committor to guide subsequent sampling \cite{jung2023machine}. Yet efficiently resolving this interdependence, while retaining access to the underlying reaction rates, remains challenging in complex molecular systems.

Here we introduce \name{}, a highly efficient approach for predicting reaction rates and characterizing transition mechanisms in complex molecular systems. Building on the stochastic optimal control formulation \cite{du2026rare}, \name{} unifies sampling and committor learning within a single control problem: the committor guides simulations toward reactive configurations, and the resulting trajectories refine both the committor and its associated control, without requiring a separate transition path sampling procedure. Across benchmark biomolecular conformational transitions, \name{} learns accurate committor functions with order-of-magnitude faster convergence. Beyond individual systems, \name{} demonstrates transferability across ion assembly systems in explicit solvent while maintaining accurate reaction rate predictions. Learning on coarse-grained coordinates further extends its reach to larger host–guest binding systems in explicit solvent. These efficiency gains also enable the study of chemical reactions using a computationally demanding machine learning interatomic potential. Together, these results establish \name{} as a basis for a foundation model to predict rare event mechanisms and kinetics across biological and chemical systems.

\section{Results}

\subsection*{Stochastic optimal control for rare event analysis}

We study transitions between two disjoint basins, $A$ and $B$, through the committor function. For a phase point $z=(x,v)$ specifying atomic positions and velocities, the committor gives the probability that the unbiased dynamics reaches $B$ before $A$:
\begin{equation}
q(z)=\Prob_z(\tau_B<\tau_A),
\label{eq:definition}
\end{equation}
where $\tau_A$ and $\tau_B$ are the first hitting times of the two basins. Values near zero indicate commitment to $A$, values near one indicate commitment to $B$, and the isosurface $q=1/2$ identifies transition states with equal probabilities of the two outcomes.

These probabilities are defined by the underlying molecular dynamics, which we model using the underdamped Langevin equation,
\begin{equation}
\dd Z_t=b(Z_t)\,\dd t+\sigma\,\dd W_t,
\qquad Z_0\sim\mu_0,
\label{eq:reference}
\end{equation}
where $Z_t=(X_t,V_t)$, $\mu_0(\dd x,\dd v)$ is the joint probability measure over positions and velocities, and $W_t$ is standard Brownian motion. The drift and noise amplitude are
$b(x,v)=\bigl(\begin{smallmatrix}
v\\
-M^{-1}\nabla U(x)-\gamma v
\end{smallmatrix}\bigr)$
and
$\sigma=\bigl(\begin{smallmatrix}
0\\
\sqrt{2\gamma k_B T M^{-1}}
\end{smallmatrix}\bigr)$, with the potential energy $U$, diagonal mass matrix $M$, friction rate $\gamma$, Boltzmann constant $k_B$, and temperature $T$.

The committor also determines how these dynamics can be guided toward a rare transition. This connection is formalized through stochastic optimal control\cite{du2026rare}, in which a control modifies the drift:
\begin{equation}
\dd Z_t^u=
\left[b(Z_t^u)+\sigma u_t(Z_t^u)\right]\dd t
+\sigma\,\dd W_t,
\qquad Z_0^u\sim\mu_0,
\label{eq:controlled}
\end{equation}
where the optimal control that ensures arrival at $B$ before $A$, which generates the transition path ensemble, is $u^*(z)=\sigma^\top\nabla_z\log q(z)$, and the flexibility to sample from arbitrary $\mu_0$ provides exploration.

\begin{figure*}[t]\centering
    \includegraphics[width=\textwidth]{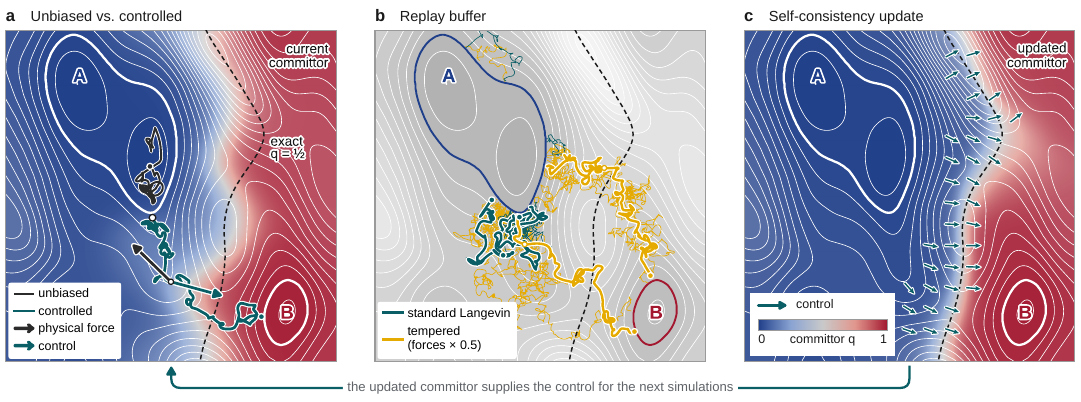}
\caption{\textbf{\name{} iterates between flexible simulation and committor learning.}
\textbf{a},~The current committor estimate defines a control that drives simulations toward otherwise rare transitions. An unbiased trajectory remains trapped in $A$, whereas controlled trajectories can cross the barrier toward $B$; arrows illustrate the physical force and the learned control.
\textbf{b},~\name{} can learn from trajectories generated under different simulation dynamics. Controlled trajectories, including simulations with modified forces or noise, are stored in a shared replay buffer. The self-consistency objective explicitly accounts for the dynamics that generated each trajectory, allowing data collected under controls in earlier iterations and sampling conditions to be reused together.
\textbf{c},~The committor is updated through the self-consistency objective using segments from the selected trajectories in the buffer. The updated committor then defines the control for subsequent simulations, closing the iterative sampling learning loop.}
\label{fig:overview}
\end{figure*}

\name{} turns this connection into an iterative cycle of simulation and learning (Fig.~\ref{fig:overview}). 
We use the current committor estimate to guide simulations, then refine the estimate using the resulting trajectories.
Trajectories are stored in a replay buffer and reused across learning updates to improve data efficiency.
Each updated estimate supplies the control for subsequent simulations.
Learning is based on a self-consistency relation for the committor. The relation includes the control used during sampling, so trajectories generated with an earlier control remain useful after the model updates. It also allows flexibility in altering the noise amplitude to encourage exploration. 
For some systems, we make learning more tractable by adopting
the commonly used approximation of the committor using
a selected set of collective variables (CVs),
$q(z)\approx q(\xi(z))$, rather than the complete state,
and considering the velocity-averaged committor $q(x)$
on atomic positions.
For example, these representations can omit solvent and
hydrogen coordinates from the inputs, while the corresponding
degrees of freedom remain explicitly represented in the
molecular dynamics simulations.
Details about the objective, replay buffer, and numerical
integration are provided in Methods.

\subsection*{Highly efficient committor learning via \name{}}

We first establish the effectiveness of Pharos on widely studied biomolecular dynamics. Alanine dipeptide, chignolin, and Trp-cage are established systems of biomolecular conformational transitions across scales, ranging from 22 atoms in a gas-phase model of alanine dipeptide to over 10,000 atoms in explicitly solvated Trp-cage, and thus serve as strong test cases for the scalability of \name{}, in comparison to the semigroup method, which relies on unbiased simulation to learn committor functions in a similar self-consistent manner\cite{strahan2021long,rotskoff2022active}. 

\begin{figure*}[!t]\centering
\includegraphics[width=\textwidth]{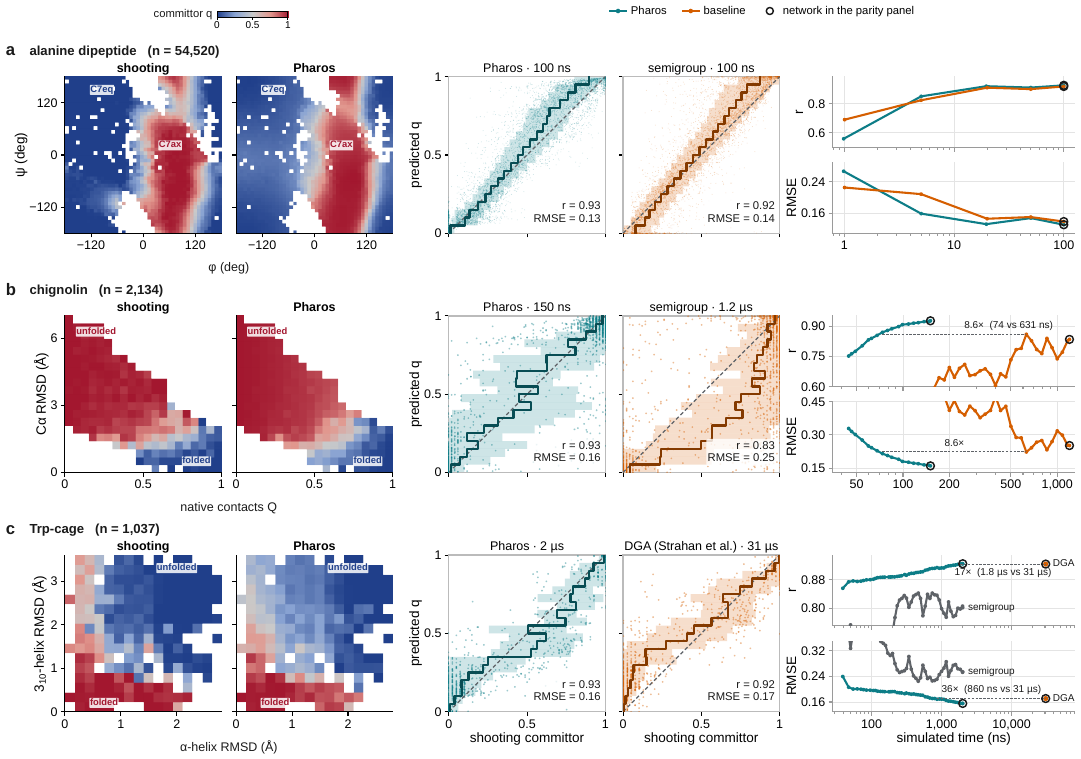}
\caption{\textbf{\name{} enables highly efficient committor learning.}
\textbf{a-c},~Alanine dipeptide, chignolin and Trp-cage.
Left, committors estimated by unbiased shooting and predicted by \name{}, projected onto the same collective variables. 
$n$ denotes the total number of test data points.
Center, predicted versus shooting committors for \name{} (teal) and the baseline (orange): semigroup learning\cite{mitchell2024committor} in \textbf{a,b} and DGA\cite{strahan2021long} in \textbf{c}.
Solid lines and shading show binned shooting means and $\pm1$ standard deviation; dashed diagonals indicate perfect agreement.
Right, Pearson correlation and RMSE versus cumulative simulated time throughout training. Teal and orange denote \name{} and the corresponding baseline; gray denotes semigroup learning on Trp-cage. Open circles mark the checkpoints shown in the center panels.}
\label{fig:si-pep-shoot}
\end{figure*}

Alanine dipeptide provides a simple model of local backbone conformational change with two possible transition paths. Due to the simplicity of the transformation, both \name{} and the semigroup method learn accurate committors within a similar simulation budget (Fig.~\ref{fig:si-pep-shoot}a). Chignolin, a minimal model of collective folding, offers a more challenging scenario in which $\beta$-hairpin formation couples local turn formation to contacts between residues distant in sequence. The semigroup method struggles to fit the committor, while \name{} stably converges to the ground truth within half the simulation time (Fig.~\ref{fig:si-pep-shoot}b). Trp-cage further couples helix formation to hydrophobic core packing, with intermediate metastable states and competing folding pathways\cite{strahan2021long}. \name{} demonstrates the advantage of controlled simulation in bypassing the energy barriers and reduces the simulation required to learn an accurate committor by a factor of 17-36 compared to the semigroup method (Fig.~\ref{fig:si-pep-shoot}c)\cite{strahan2021long}. Overall, these benchmarks demonstrate the significant computational efficiency enabled by \name{}, thanks to its flexibility in exploring the potential energy surface with controlled simulation.

\subsection*{\name{} identifies kinetic barriers in host-guest binding}

\begin{figure}[t]
    \centering
    \includegraphics[width=1\linewidth]{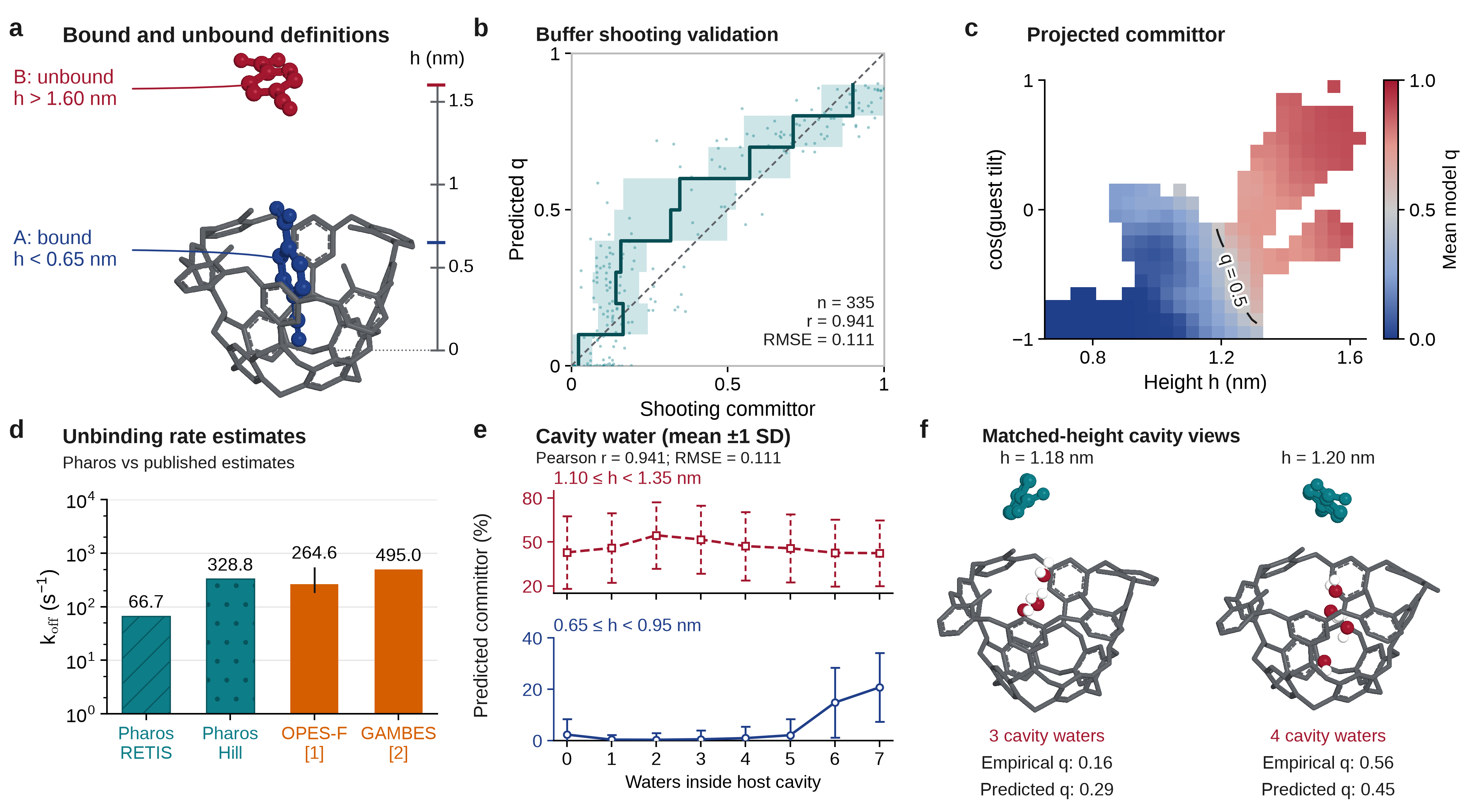}
\caption{\textbf{\name{} learns an explicit-solvent committor for host-guest binding.}
\textbf{a},~Bound and unbound reference structures and basin definitions.
\textbf{b},~Predicted versus shooting committors; lines and shading show binned shooting means and $\pm1$ standard deviation.
\textbf{c},~Predictions projected onto guest height and tilt; purple curve, projected $q=0.5$ contour.
\textbf{d},~Preliminary unbinding-rate estimates from RETIS and the Hill relation, compared with published computational estimates.
\textbf{e},~Predicted committor versus cavity water count for guests near the host (blue) and farther away (red); error bars, $\pm1$ standard deviation.
\textbf{f},~Two configurations at similar guest heights but with different shooting committors, alongside the model predictions.
}
\label{fig:host_guest}
\end{figure}

Host-guest systems have applications in drug discovery~\cite{braegelman_integrating_2019} and materials science~\cite{zhu_applications_2023}, and provide a tractable setting for studying rare binding events. Despite their small size, these systems can exhibit dissociation timescales extending to milliseconds~\cite{debnath_computing_2021,thomas_tuning_2017}, making their kinetics challenging to characterize through direct simulation. Their extensive use in the SAMPL challenges has established them as well-studied benchmarks for computational models of molecular recognition~\cite{yin_overview_2017,amezcua2024sampl9}. This combination of computational tractability, slow dissociation, and extensive prior characterization makes host–guest systems valuable tests of methods for predicting binding kinetics, with potential applications in drug discovery and materials design.

We successfully tackle these challenges in \name{}. Simulations were performed in explicit solvent using the reference system and dynamics established by the SAMPL Challenge \cite{yin_overview_2017,rizzi_overview_2018}. We parameterized the committor using an equivariant graph neural network \cite{painn2021} that incorporates the positions and momenta of the host, guest, nearby water oxygen atoms, and ions. The committor predicted by \name{} shows excellent agreement with long unbiased MD shooting simulations, achieving a Pearson correlation coefficient of $r = 0.94$ across 335 test configurations (\Cref{fig:host_guest}b). These results demonstrate that \name{} can accurately learn the committor for processes occurring on microsecond timescales. Furthermore, we show that the committor depends not only on the host–guest separation but also on factors such as guest orientation and cavity water occupancy (\Cref{fig:host_guest}c). As shown in \Cref{fig:host_guest}d, \name{} captures the strong influence of packed cavity waters on binding when the guest is close to the host, while correctly identifying that this influence becomes negligible as the guest moves farther away. \Cref{fig:host_guest}f illustrates two configurations in which the guests occupy similar heights but differ in orientation and cavity water occupancy, leading to distinct unbinding probabilities that \name{} accurately predicts. Finally, we use \name{} to calculate the kinetic $k_{off}$ rate. Using both the hill equation \Cref{eq:si-flux}, and replica exchange transition interface (RETIS)\cite{vanerp2007retis}, we discover that our rates are within order of magnitude to published baselines \cite{debnath_computing_2021,thomas_tuning_2017}.

\subsection*{\name{} enables transferable committor learning and reveals solvent effect of ion pair dissociation}

\begin{figure*}[t]\centering
\includegraphics[width=\textwidth]{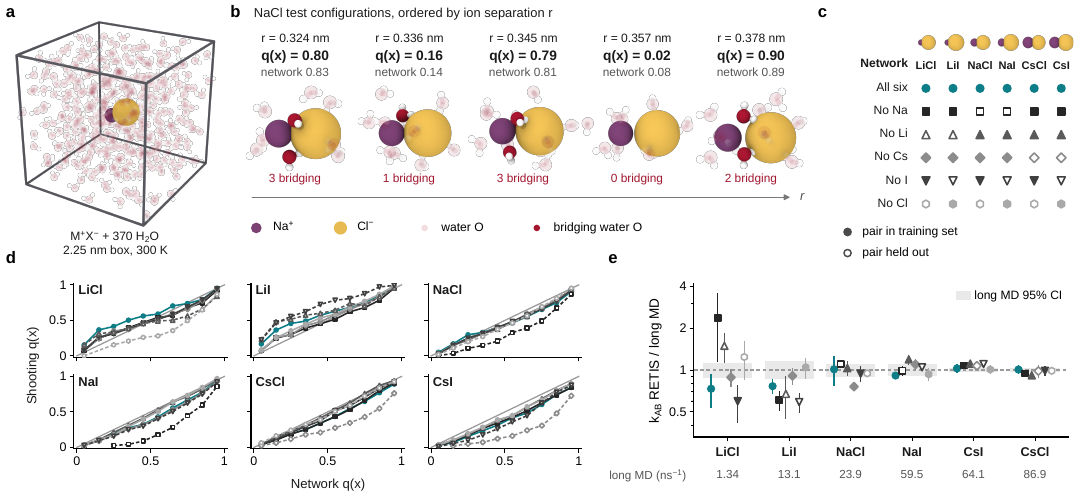}
\caption{\textbf{A learned committor transfers across ion pairs and gives accurate dissociation rates.}
\textbf{a},~The simulation box: one cation-anion pair in water.
\textbf{b},~Five NaCl test configurations ordered by ion separation $r$, with their shooting committors $q(x)$ and the predictions of the
network trained on all six pairs. The committor follows the number of bridging waters (opaque; other waters translucent).
\textbf{c},~The six networks and the pairs each was trained on (filled) or never saw (open); teal, the network trained on all six pairs; grays,
networks trained without one ion.
\textbf{d},~Shooting committor against each network's $q(x)$, 1,200 configurations per pair (mean in 0.1-wide bins; markers and colors as in
\textbf{c}; dashed, held-out pair; thin gray line, identity).
\textbf{e},~Rate constants $k_{AB}$ from the contact pair $A$ to the separated pair $B$, from  {RETIS}  {using}  {interfaces}  {defined}  {by} each network's $q_\theta(x)$, relative to long plain MD (gray band, 95\% interval); mean $\pm$ s.e.\ of four runs; pairs ordered by the
long MD rate constant, shown below the plot in gray.}
\label{fig:ions}
\end{figure*}

Ion pair dissociation in solvent is an important precursor to understanding many biological and interfacial processes.\cite{van2016water}
Although ion separation provides a natural reaction coordinate, dissociation also involves collective solvent rearrangements.\cite{jung2023machine}
We use Pharos to learn a shared committor conditioned on the physical properties of the ions. We studied six salts formed by Li$^+$, Na$^+$, and Cs$^+$ with Cl$^-$ and I$^-$, each solvated by 370 explicit water molecules (Fig.~\ref{fig:ions}a). 
The model describes ions through their charges, Lennard-Jones sizes and well depths, and masses, without species-specific embeddings. 
An equivariant graph neural network~\cite{painn2021} processes all-atom configurations and velocities without predefined solvent descriptors~\cite{jung2023machine,okada2026deep}.
Training also includes seven virtual ion pairs, extending the physical parameter combinations beyond those supplied by the real ions.

First, the estimated committor from \name{} agrees well with committors estimated from independent unbiased trajectories across all six salts (Fig.~\ref{fig:ions}d). 
The model also distinguishes configurations with similar ion separations but different solvent environments. 
For example, two NaCl configurations with separations near 0.35 nm have committors of 0.79 and 0.02, compared with model predictions of 0.81 and 0.08. The former contains three waters bridging the ions, whereas the latter contains none (Fig.~\ref{fig:ions}b). Pharos captures this distinction directly from the atomic representation.

Second, \name{} provides insight into the reaction mechanism
through perturbation analysis.
From Li$^+$ to Cs$^+$, the total committor response to thermal
motion near the transition state decreases, while water
accounts for a larger fraction.
Within the water response, the contribution of outer water
increases relative to that of first-shell water
(Supplementary Fig.~\ref{fig:ion-mechanism}a,b).
The first-shell total response weakens even though the number of first-shell waters
increases (Supplementary Fig.~\ref{fig:ion-mechanism}c).
To distinguish the effect of changing ion parameters from
that of changing configurations, we compare the relative
contribution of outer water under different ion parameters
at fixed configurations
(Supplementary Fig.~\ref{fig:ion-mechanism}d).
From LiCl to CsCl, both changes in the sampled configurations
and changes in ion parameters at the same configurations
increase the relative contribution of outer water.

Third, \name{} uses ion physical properties as inputs instead of the ion category, allowing transfer to ion species absent from training.
In addition to the full model, we train five models, each excluding one cation or anion and all training systems containing that ion (Fig.~\ref{fig:ions}c). Without further training on the excluded systems, their predictions remain correlated with independently estimated committors and distinguish configurations at different stages of reaction progress (Fig.~\ref{fig:ions}d).

Fourth, the learned committor acts like a reaction coordinate to support quantitative kinetic calculations. 
Reference dissociation rates across the six salts span approximately 65-fold.
Using the fixed committor models to define interfaces for replica exchange transition interface sampling (RETIS) recovers these rates with median absolute relative errors within $10\%$ for systems both represented during training and for withheld systems (Fig.~\ref{fig:ions}e). 
Pharos thus provides a transferable description of solvent response and reaction progress, enabling committor prediction and accurate dissociation-rate calculations, even for unseen ion species.

\subsection*{\name{} resolves the first step of prebiotic chemistry in hydrogen cyanide}

\begin{figure*}[t]\centering
    \includegraphics[width=\textwidth]{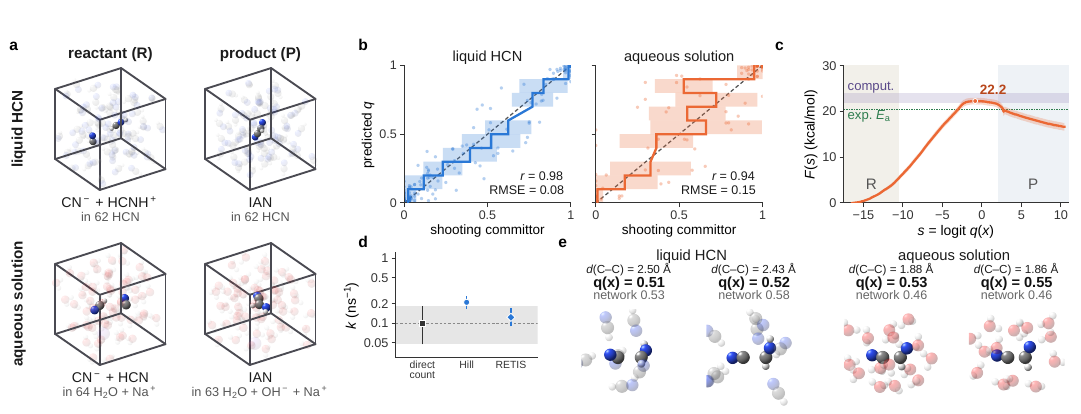}
    \caption{\textbf{\name{} resolves the dimerization of hydrogen cyanide in liquid HCN and in aqueous solution.}
    \textbf{a},~Reactant and product boxes in liquid HCN and aqueous solution phases. Reactive species are opaque, and the solvent is transparent.
    \textbf{b},~Predicted versus shooting committors for test configurations. Solid lines and shading show binned shooting means and $\pm1$ standard deviation; dashed diagonals indicate perfect agreement.
    \textbf{c},~Free energy along $s=\operatorname{logit}q$ from umbrella sampling in aqueous solution. Shading marks $\pm2$ s.e.\ from bootstrap over replicas. R and P mark the reactant and product states. The purple band stands for computed barriers for cyanide adding to HCN, from 21.8\,kcal\,mol$^{-1}$
    ~\cite{sandstromBeginningHCNPolymerization2021} to 24.0\,kcal\,mol$^{-1}$~\cite{cappellettiHowDoesAdenine2026}. The green dotted line stands for the experimental activation energy of 20.4\,kcal\,mol$^{-1}$
    ~\cite{sanchezStudiesPrebiodcSynthesis1967}.
    \textbf{d},~Rate constants of the ion pair channel CN$^-$ + HCNH$^+$ $\to$ IAN in liquid HCN: direct counting in plain MD, the Hill relation using $q(x)$, and RETIS using interfaces defined by $q(x)$. The bars stand for 95\% intervals.
    \textbf{e},~Structures in the transition state ensemble near the $q=1/2$ surface, with the length of the forming C-C bond, the shooting committor $q(x)$ and the network prediction. Reacting molecules are opaque and nearby solvent molecules are transparent.}
    \label{fig:reaction}
\end{figure*}

Hydrogen cyanide (HCN) is one of the most widespread molecules in the universe, and was likely prevalent in the atmosphere of the early Earth~\cite{sandstromCrossroadsOriginPrebiotic2023}. Since 1960, when Or\'o showed that concentrated ammonium cyanide
solutions yield adenine~\cite{oroSynthesisAdenineAmmonium1960}, the building block of DNA and RNA, HCN has been placed at the root of prebiotic reaction networks that lead to the origin of life~\cite{patelCommonOriginsRNA2015}, however, how HCN assembles into adenine formally through \ce{5 HCN -> C5H5N5} remains unresolved in the past decades. All proposed routes start from the same transformation: formation of
the HCN dimer iminoacetonitrile (IAN)~\cite{cappellettiHowDoesAdenine2026}. This dimerization is a rate-determining step of the entire prebiotic HCN self-reaction network, yet is hard to study experimentally or computationally. Experiments estimate the apparent rate by the downstream products but are not well-suited to capture the transient IAN species from the complex reaction mixture~\cite{ruiz2021comprehensive}, while computational studies often have to use static structures and simplify the solvent environment in quantum chemistry cluster studies~\cite{cappellettiHowDoesAdenine2026}, or rely on a hand-picked reaction coordinate and suffer from the prohibitive cost in \textit{ab initio} molecular dynamics~\cite{sandstromBeginningHCNPolymerization2021}. 

We study this reaction with \name{} using UMA~\cite{woodUMAFamilyUniversal2025,levineOpenMolecules20252026}, one of the most advanced machine learning interatomic potentials (MLIPs), which enables studying chemical reactions at quantum-chemistry accuracy with nanosecond-scale simulations. Nevertheless, the quantitative characterization of chemical reactions remains computationally hard due to the prohibitive cost of committor analysis in previous work. For example, recent transition path sampling studies of reactions in water with MLIPs relied on extensive transition path ensembles, often
amounting to hundreds of nanoseconds of simulation, to characterize their transition states~\cite{davidCompetingReactionMechanisms2024,garciamartinezProtonationMagnesiumIons2026}.
By concentrating sampling in the transition region through controlled simulations, \name{} learns a committor from shorter simulations, on the order of nanoseconds for each system studied here, about two orders of magnitude less than those path-sampling studies, and turns them into effective descriptions of reactivity. For IAN formation in two condensed phases (Fig.~\ref{fig:reaction}a; Supplementary Section~\ref{sec:si-sys-hcn}): HCN-autoionization species \ce{CN-} + \ce{HCNH+} in the liquid HCN, and \ce{CN- (aq)} + \ce{HCN (aq)} as in the experimental study~\cite{sanchezStudiesPrebiodcSynthesis1967}, 
$q$ learned for each phase reproduced the shooting
probability of commitment with high correlation across different extents of reactions and solvent configurations (Fig.~\ref{fig:reaction}b). 

We then study how the reactivity of the species in the two phases differs, starting from their free energy barriers. Using $s=\operatorname{logit}q$ as the collective variable for umbrella sampling
removes the need for a hand-picked collective variable for free energy calculation. In aqueous solution, the barrier of 22.2\,kcal\,mol$^{-1}$ lies within 2\,kcal\,mol$^{-1}$ of the experimentally estimated
activation energy from the downstream reactions~\cite{sanchezStudiesPrebiodcSynthesis1967} and of previous
calculations~\cite{sandstromBeginningHCNPolymerization2021,cappellettiHowDoesAdenine2026} (Fig.~\ref{fig:reaction}c). This experimental barrier was associated with second-order reaction kinetics, whose measured
rate constant is $2.7\times10^{-7}$\,M$^{-1}$\,s$^{-1}$. For one cyanide at the HCN
concentration of our box (0.9\,M), this corresponds to $2.4\times10^{-7}$\,s$^{-1}$, i.e., about one reaction every seven weeks~\cite{sanchezStudiesPrebiodcSynthesis1967}. In liquid HCN, we computed the rate of IAN formation from the \ce{CN-} + \ce{HCNH+} ion pair directly. The ion pair forms IAN with a rate constant of about $10^{8}$\,s$^{-1}$, i.e., one event within about 10\,ns. Therefore, the ion pair reacts some fourteen orders of magnitude faster, confirming its exceptional reactivity. Compared to the rate directly counted from MD trajectories, our rates derived from $q$ agree within a factor of two and reduce the required length of trajectory from 100\,ns to 10\,ns with similar statistical uncertainty (Fig.~\ref{fig:reaction}d).

To understand the origin of this reactivity difference, we examine the transition state ensemble defined by the $q=1/2$ surface
without any assumption about its mechanism (Fig.~\ref{fig:reaction}e). Among those structures, the two reacting carbons are about 0.5\,\AA\ closer in aqueous solution than in liquid HCN. In liquid HCN, every transition state is a \ce{CN-} facing \ce{HCNH+}: the ion pair commits to the dimer earlier, driven by the Coulomb attraction between the opposite charges, the strongly electrophilic nitrilium carbon. In aqueous solution, the anion must attack a neutral HCN and push its charge onto the
imine nitrogen of \ce{NC-CH=N-}, so the transition state is late and close to
the product. An early transition state with a low barrier, in contrast with a late one with a high barrier, is consistent with the well-known Hammond
postulate~\cite{hammond1955correlation}, here resolved from dynamics in explicit solvent rather than assumed from static
structures.

\section{Discussion}

In this paper, we introduced \name{}, a method that learns the committor for rare transitions from controlled simulations. One key advantage of \name{} is that the estimated committor during the training process further guides the next round of simulations toward the transition region to push the convergence of the algorithm. 
Our learning objective allows \name{} flexible sampling dynamics to encourage transitions and to combine trajectories from different sampling conditions to learn the committor in a principled manner. 
\name{} hence decouples the time needed to sample rare transitions from their residence waiting times, making it possible to learn committors for events beyond the reach of direct simulation.

By learning how molecular configurations influence transition probabilities, \name{} captures coupled structural and solvent rearrangements that simple geometric reaction coordinates cannot resolve. The same learned function identifies transition state configurations and provides a reaction coordinate for free energy and reaction rate calculations, linking microscopic changes to their consequences for reactivity. This combination is particularly valuable for chemical reactions described by computationally demanding interatomic potentials, where the cost of sampling limits access to dynamical information. In the HCN systems, efficient committor learning enables the characterization of bond formation in explicit solvent and reveals how the chemical environment influences reactivity. Beyond individual systems, conditioning on physical properties allows the learned description of reaction progress to transfer to unseen ion species, supporting quantitative kinetic calculations without retraining the committor. \name{} thus provides an integrated approach to identifying what controls rare transitions and determining how frequently they occur.

One limitation of \name{} is its Markovian modeling assumption, which neglects memory effects. Although representing the committor in terms of CVs enables \name{} to scale to realistically sized systems, the resulting reduced dynamics can exhibit memory effects that are important in many biological systems~\cite{dalton2023fast}. Another challenge is ensuring adequate sampling coverage, both in the initial conditions and along simulated trajectories. Although \name{} accommodates arbitrary initial distributions, we do not explicitly investigate strategies for exploring these distributions here. Developing more efficient exploration strategies is therefore a promising direction for future work~\cite{tang2026breaking}. While \name{} provides a general framework that accommodates sampling from arbitrary initial distributions, we do not explicitly investigate strategies for exploring these distributions here.

The substantial gains in sampling efficiency enabled by \name{} open several directions for future work. Transfer across ion pairs suggests a route toward transferable foundation models for rare event analysis. A model pretrained across molecular systems and transitions could provide an initial committor estimate for a new transition, reducing the simulation required to identify pathways and estimate transition rates. Another promising direction is to extend \name{} to complex reaction networks described by expressive machine learning interatomic potentials, where the computational cost of long-timescale simulations remains a major barrier. More broadly, extending \name{} to collective transformations in condensed phase systems could advance our understanding of phenomena such as crystallization.

\section{Methods}\label{sec:methods}
\subsection*{Learning \name{} with control}

\paragraph{Committor and the backward Kolmogorov equation.} The committor (Eq.~\eqref{eq:definition}) is known as the solution of the backward Kolmogorov equation (BKE)
\begin{equation}
\Lop q=0\quad\text{outside }A\cup B,\qquad q|_A=0,\quad q|_B=1,
\label{eq:q}
\end{equation} 
where $\Lop$ is the generator of the reference dynamics in Eq.~\eqref{eq:reference}, which is defined as the expected instantaneous rate of change, i.e., $\Lop q(z)=\lim_{t\to0}(\E_z[q(Z_t)]-q(z))/t$, and takes the form 
\begin{equation}
\Lop q=b\cdot\nabla q+\tfrac12\sigma\sigma^\top:\nabla^2q.
\label{eq:generator}
\end{equation}

\paragraph{The \name{} objective.}
\name{} leverages the stochastic optimal control formulation, leveraging the BKE formulation from the optimal controlled process, where $q$ is known as the Doob's h-transform\cite{jamison1975markov}. \name{} minimizes the squared residual of BKE as in physics-informed neural networks\cite{raissi2019physics},
\begin{align}
\label{eq:J_initial}
\mathcal J(\theta)&=\E_{\mathcal D}\left[\left|\int_{t_1}^{t_2} \kappa_s\,\Lop q_\theta (Z_s) \, \dd s \right|^2\right],
\end{align}
where $\mathcal D$ includes trajectories sampled from arbitrary dynamics $\dd Z_t=\widetilde b_t\,\dd t+\sqrt{\kappa_t}\,\sigma\,\dd W_t$, with sampling drift $\widetilde b_t$ and noise multiplier $\kappa_t>0$, and $t_1 < t_2$ are arbitrary times. Setting $\widetilde b_t=b$ and $\kappa_t=1$ recovers the reference dynamics. The noise multiplier scales the variance of the thermal noise; for example, $\kappa_t>1$ with the physical drift unchanged ($\widetilde b_t=b$, possibly plus a control) amounts to running the thermostat at the higher temperature $\kappa_tT$. To avoid the expensive calculation of the Laplacian term in $\Lop q_\theta$, we use It\^o's lemma to rewrite the integral of the residual in terms of $q$ and $\nabla q$ alone. In the end, the objective becomes
\begin{align}
\label{eq:J_final}
\mathcal J(\theta)&=\E_{\mathcal D}\left[\left|q_{\bar\theta}(Z_{t_2})-q_\theta(Z_{t_1}) - \int_{t_1}^{t_2} (\widetilde b_t(Z_t)- \kappa_t b(Z_t))\cdot\nabla q_\theta(Z_t)\,\dd t - \int_{t_1}^{t_2}\sqrt{\kappa_t}\,\nabla q_\theta(Z_t)^\top\sigma\,\dd W_t \right|^2\right],
\end{align}
A natural choice is the controlled drift $\tilde{b}_t = b + \sigma u$ introduced in \Cref{eq:controlled}, with $u(z) = \sigma^{\top} \nabla_z \log q_{\theta}(z)$. In practice, we further alter the sampling dynamics to encourage exploration, as discussed below.

Supplementary Section~\ref{sec:si-methods} provides the derivation and analysis of the objective. The benefit of \name{} objective comes from its flexible sampling dynamics, off-policy training, and tractable evaluation.

\paragraph{Sampling dynamics and replay.}

Training of \name{} alternates between simulation and committor updates.
We start simulations from an initial set of configurations or from configurations saved during earlier simulations.
By default, each simulation targets basin $B$ with probability given by the learned committor and basin $A$ otherwise.
We simulate the controlled dynamics in Eq.~\ref{eq:controlled}.
For target $B$, the control uses the gradient of $\log q$; for target $A$, it uses the gradient of $\log(1-q)$.
This control changes the velocities only to favor reaching the selected basin.
For some systems, we also scale the physical forces, change the simulation temperature 
and/or alter selected interactions to encourage
transitions.
Each simulation runs for a prescribed duration
and stops early if it enters either basin.
We store trajectories in a replay buffer, together with
the applied control, force modifications, and noise increments.
This information allows us to evaluate the loss on segments selected from the stored trajectories.

Each network update selects segments from the stored
trajectories and evaluates the self-consistency loss
(Eq.~\ref{eq:J_final}).
Since the self-consistency relation holds on each segment, a single trajectory can provide training signals from different starting points and over different durations.
Depending on the system, we use fixed or randomly chosen
starting points of preferentially select segments
ending at basin entry. For segments ending at basin entry, the known basin value of the committor replaces the predicted endpoint value.
Since this loss accounts for the dynamics that generate each segment, it allows using trajectories from earlier controls and different sampling conditions to be reused while learning the committor of the reference dynamics.
The updated committor then guides subsequent simulations.
Starting initial configurations, force modifications,
trajectory lengths, buffer sizes, and segment selection
are specified in Supplementary Section~\ref{sec:si-exp}.

\subsection*{Molecular representations and systems}

\paragraph{Alanine dipeptide.}
Alanine dipeptide contains 22 atoms and is simulated in vacuum
with CHARMM27 at 300\,K.
The reactant and product basins are centered at
$(\phi,\psi)=(-150^\circ,170^\circ)$ and
$(90^\circ,-50^\circ)$, corresponding to C7eq and C7ax.
Each basin includes configurations within $10^\circ$ of its
center in both dihedral angles.
We use a multilayer perceptron (MLP) to learn the
phase-space committor $q(x,v)$.
Positions enter through all 45 pairwise distances among
the ten heavy atoms.
Velocities enter through the rates of change of these distances.

\paragraph{Protein folding.}
Chignolin (CLN025) is a ten-residue protein simulated with
CHARMM27 in 2,100 TIP3P water molecules with two sodium
counterions at 340\,K.
The folded and unfolded basins are defined by $Q\geq0.99$
and $Q\leq0.01$, respectively, where $Q$ is the fraction
of native contacts.
We learn $q(x,v)$ with an MLP using the 93 protein
heavy atoms. Positions are represented by all pairwise
distances between these atoms. Velocities are represented
by the rates of change of these distances, calculated
from the relative atomic velocities.

Trp-cage (TC5b) is a 20-residue protein simulated with
CHARMM36m in 3,732 TIP3P water molecules, with ten K$^+$
and eleven Cl$^-$ ions, at
300\,K\cite{strahan2021long,charmm36m}.
The folded and unfolded basins follow the reference
study's joint criteria on the C$\alpha$ RMSDs of the
$\alpha$-helix (residues 2-9) and the $3_{10}$-helix
(residues 11-15), measured relative to the folded structure.
The folded basin additionally requires a distance below
1.7\,nm between the first and last C$\alpha$ atoms.
The precise RMSD boundaries are given in the Supplementary
Information.
We use a PaiNN-type equivariant graph neural network
\cite{painn2021} to approximate the committor $q(x,v)$.
The network takes the positions and velocities of the
20 C$\alpha$ atoms, together with their residue identities,
as inputs.
Both protein simulations retain all protein and solvent
atoms, although the networks use only protein information.

\paragraph{Ion pairs and transfer across species.}
Each ion system contains one ion pair and 370 TIP3P water
molecules at 300\,K, with Joung-Cheatham ion
parameters\cite{joung2008,jorgensen1983}.
The six salts combine Li$^+$, Na$^+$ and Cs$^+$ with
Cl$^-$ and I$^-$.
Contact and dissociated basins are defined by ion separation,
using the thresholds in
Supplementary Table~\ref{tab:ion pairs}.
We use a PaiNN-type equivariant graph neural network
\cite{painn2021} to approximate the committor $q(x,v)$
from the positions and velocities of both ions and
all water atoms.
The network also receives relative velocity information
and Coulomb and Lennard-Jones interaction energies between
neighboring atoms.
Each ion is specified by its charge, mass, and Lennard-Jones parameters only instead of a species label.

We train one model on all six salts and five additional
models, each excluding every salt containing one selected
ion species.
Training also draws on a set of seven virtual salts with
varied ion sizes, interaction strengths and masses.
Virtual salts containing an excluded physical ion are
withheld from the corresponding model.
Predictions for excluded salts use the trained networks
without further fitting.
The construction and selection of virtual salts are
described in the Supplementary Information.

\paragraph{Host-guest binding.}
The host-guest system contains the OAMe host and G2 guest
(4-cyanobenzoate), 2,100 TIP3P water molecules and nine
sodium counterions at 300\,K.
Simulations use the reference GAFF/RESP parameters and
funnel restraint\cite{rizzi_role_2021}.
Bound and dissociated basins are defined by the guest
height $h$ along the host axis:
$0<h<0.65$\,nm and $h>1.60$\,nm, respectively.
We use a PaiNN-type network\cite{painn2021} to approximate $q(x,v)$
from the positions and velocities of the 143 solute
heavy atoms, nearby water oxygens and sodium ions.

\paragraph{Hydrogen cyanide systems.}
Simulations of both HCN systems used the UMA (OMol task) MLIP\cite{woodUMAFamilyUniversal2025,levineOpenMolecules20252026} with Langevin dynamics at 300\,K. Liquid HCN contains 64 HCN units, and the aqueous solution contains one CN$^-$, one HCN, and one Na$^+$ in 64 waters. States are defined on the molecular graph: the product is any C--C bond of at most 1.60\,\AA; the reactant is a free \ce{CN-} + \ce{HCNH+} ion pair or the neutral liquid in liquid HCN, and a C--C distance of at least 3.00\,\AA\ in the aqueous solution. The network is a PaiNN-type encoder\cite{painn2021} on atomic positions and velocities whose readout pools over all carbon pairs. Training configurations come from independent steered MD trajectories, and test configurations come from separate trajectories not used in training (Supplementary Section~\ref{sec:si-sys-hcn}).

Further details of each system are provided in
Supplementary Section~\ref{sec:si-systems}.
The networks used to model each system are described in
Supplementary {Section}~\ref{sec:si-networks}.

\subsection*{Evaluation}

\paragraph{Committor accuracy and sampling efficiency.}
We compare the learned committor with estimates from
unbiased shooting under the reference dynamics.
From each test configuration, we launch independent trajectories
with thermal velocities and record which basin they reach first.
The shooting estimate is the fraction reaching $B$
among trajectories that reach either basin.
For comparison, we average the network predictions over velocities:
\begin{equation}
q_\theta(x)=
\mathbb{E}_{v\sim p_{\mathrm{MB}}(\cdot\mid x)}
\left[q_\theta(x,v)\right],
\end{equation}
where $p_{\mathrm{MB}}$ is the Maxwell-Boltzmann distribution
at the system temperature, subject to the molecular constraints.

We measure agreement using Pearson correlation and
root-mean-square error over the test configurations.
For alanine dipeptide and the proteins, we compare methods
on the same configurations and track accuracy against
total simulated time, including simulations used to
generate the initial configurations.
For ion transfer, the full and excluded-ion models use
the same test sets, without further training on excluded salts.
Test-set construction and shooting procedures are given in
Supplementary Section~\ref{sec:si-eval}.

\paragraph{Reaction rates.}
We use the learned committor to calculate rates with
replica exchange transition interface sampling
(RETIS)~\cite{vanerp2007retis}
and the Hill relation~\cite{kramers1940brownian,baudel2023hill}.

For RETIS, we place interfaces at selected committor values
between $A$ and $B$.
We sample trajectories crossing each interface and estimate
the fraction that reaches the next interface before returning to $A$.
Multiplying these fractions by the exit flux from $A$
gives the transition rate.

For the Hill relation, we estimate the rate directly
from simulation and the learned committor.
After each departure from $A$, we record the first crossing
of a chosen interface before the trajectory returns to $A$
or reaches $B$.
We weight each crossing by its predicted committor
of reaching $B$ before $A$.
The sum of these weights estimates the number of transitions
from $A$ to $B$.
We obtain the rate by dividing this sum by the total time
spent in $A$, plus the time after each departure from $A$
until either basin is reached.

The interface settings, simulation lengths, and uncertainty
estimates are given in Supplementary Section~\ref{sec:si-eval}.

For further details on committor accuracy, rate calculations, and more system-specific evaluation procedures, see Supplementary Section~\ref{sec:si-eval}.

\section*{Acknowledgments}
We thank Peter Bolhuis, Jutta Rogal, Jonathan Weare and Chatipat Lorpaiboon for useful discussions and pointers to relevant literature. 
JH acknowledges support from the University of Cambridge Harding Distinguished Postgraduate Scholars Programme. WL and HJK acknowledge support from the Wang Family in the form of a Faculty Research Innovation Fund.

\clearpage

\providecommand{\Lop}{\mathcal{L}}
\providecommand{\Prob}{\mathbb{P}}
\providecommand{\E}{\mathbb{E}}
\providecommand{\dd}{\mathrm{d}}
\providecommand{\Pst}{P^{\mathrm{stop}}_T}

\providecommand{\Dom}{\Omega}
\providecommand{\bt}{\widetilde b}
\ifcsname sprop\endcsname\else
  \newtheorem{sprop}{Proposition}
  \renewcommand{\thesprop}{S\arabic{sprop}}
  \theoremstyle{remark}
  \newtheorem{srem}{Remark}
  \renewcommand{\thesrem}{S\arabic{srem}}
\fi
\setcounter{section}{0}
\renewcommand{\thesection}{S\arabic{section}}
\setcounter{equation}{0}
\renewcommand{\theequation}{S\arabic{equation}}
\setcounter{figure}{0}
\renewcommand{\thefigure}{S\arabic{figure}}
\setcounter{table}{0}
\renewcommand{\thetable}{S\arabic{table}}

\clearpage
\pdfbookmark[0]{Supplementary Information}{supplementary}
\begin{center}
  {\fontsize{24}{28}\selectfont\bfseries Supplementary Information}
\end{center}
\vspace{1em}
\startcontents[sections]
\printcontents[sections]{l}{1}{\setcounter{tocdepth}{3}}

\titlecontents{lsubsection}[3em]
  {}                                  
  {\contentslabel{2em}}              
  {}                                  
  {\titlerule*[.6em]{.}\contentspage}  

\clearpage

\section{\name{} method}
\label{sec:si-methods}

Section~\ref{sec:si-prior} reviews the self-consistency objectives for the
committor used in previous work.
Section~\ref{sec:si-ours} relates the \name{} objective to them and provides the first-order optimality guarantee and the convergence rate.

\subsection{Previous self-consistency objectives}
\label{sec:si-prior}

We can express the existing self-consistency objectives for committor learning in terms of the stopped transition operator. Given the process $Z_t$ defined in equation \eqref{eq:reference}, and its hitting time  on $A \cup B$, $\tau := \tau_A \wedge \tau_B$, the stopped transition operator is defined as
\begin{equation}
\Pst q(z)=\E_z\big[q(Z_{\tau\wedge T})\big].
\label{eq:si-stopped}
\end{equation}
By construction, the committor $q$ is characterized as the function such that $\Lop q=0$ in $\mathbb{R}^d \setminus (A \cup B)$, $q=0$ in $A$ and $q=1$ in $B$, and the condition $\Lop q=0$ is equivalent to $\Pst q=q$ (Proposition 1, Eq. (9) of Li et al., 2022\cite{li2022semigroup}). 
All methods below turn the identity $\Pst q=q$ into a loss; they differ in
the estimator, in whether the target is detached ($\bar \theta$ denotes
detached neural network weights) and in whether the terminal state is stopped at $A\cup B$.
Trajectories follow the reference dynamics and start from $Z_0\sim\mu_0$.

\begin{itemize}
\item[(i)] \textbf{Semigroup loss} \cite{li2022semigroup,strahan2023predicting}.
\begin{equation}
\mathcal J_{\mathrm{SG}}(\theta)=\E\big[(\Pst q_{\theta}-q_{\theta})^2(Z_0)\big].
\label{eq:si-LSG}
\end{equation}
The gradient is a product of two conditional expectations, so an unbiased
estimate needs at least two independent trajectories per initial point.
Nothing is detached, and the trajectories are stopped at $A\cup B$.
\item[(ii)] \textbf{Logarithmic semigroup loss} \cite{mitchell2024committor}.
\begin{equation}
\mathcal J_{\mathrm{LOG}}(\theta)=\E\big[(\log q_{\theta} -\log\Pst q_{\bar \theta})^2(Z_0)\big],
\label{eq:si-LLOG}
\end{equation}
plus the same term for $1-q_{\theta}$; $\Pst\bar q_{\bar \theta}(Z_0)$ is estimated with a
detached set of $k$ trajectories ($k=100$). Because the logarithm penalizes relative errors, the loss stays sensitive where $q_{\theta}$ or $1-q_{\theta}$ is close to zero, i.e. near the basins. (Their implementation stops all $k$ trajectories as soon as one of them enters $A\cup B$. This common time is still a stopping time for each trajectory; by the optional stopping theorem the average remains an unbiased estimate of $q(Z_0)$ at $q_{\theta}=q$.)
\item[(iii)] \textbf{Variational committor network (VCN)}
\cite{he2022committor,chen2023discovering,tang2026breaking}.
\begin{equation}
\mathcal J_{\mathrm{VCN}}(\theta)=\E_\pi\big[(q_{\theta}(Z_T)-q_{\theta}(Z_0))^2\big],
\label{eq:si-LVCN}
\end{equation}
averaged over lag-$T$ pairs along equilibrium trajectories ($Z_0\sim\pi$,
the invariant law), with the boundary values imposed on $f$. The terminal
state is \emph{not} stopped and both endpoints are differentiated. At
$q_{\theta}=q$ and for reversible dynamics, $\mathcal L_{\mathrm{VCN}}/2T$ is the
reactive flux, which motivates its minimization.

\end{itemize}

We say that loss is \emph{first-order
optimal} if $q$ is its only stationary point, i.e.\ the only $q$ at which
the directional derivative vanishes for every $\eta$; for stop-gradient
updates the same notion is that $q$ is the unique \emph{fixed point}.

\begin{srem}[Optimality of previous objectives] The semigroup and logarithmic semigroup losses are first order optimal, while the VCN loss is not in general:
\label{rem:si-prior}
\begin{enumerate}[label=(\roman*), leftmargin=*]
\item \emph{Semigroup loss.} $\mathcal J_{\mathrm{SG}}$ is a convex
quadratic in $q_{\theta}$ that vanishes exactly when $\Pst q_{\theta}=q_{\theta}$ $\mu$-a.e.; since $q$
is the only such function with the prescribed boundary values,
$\mathcal J_{\mathrm{SG}}$ is first-order optimal
\cite[Prop.~E.1]{du2026rare}.
\item \emph{Logarithmic semigroup loss.} With the target detached, the
update vanishes in every direction iff $\Pst q_{\theta}=q_{\theta}$, so $q$ is its unique
fixed point \cite[Thm.~E.2]{du2026rare}. This holds for the exact target
$\log\Pst q_{\bar \theta}$; if $\Pst q_{\bar \theta}$ is replaced by a $k$-trajectory average
inside the logarithm, Jensen's inequality biases the fixed point downwards
by a relative amount of order $\operatorname{Var}(\Pst q_{\theta}(Z_{\tau\wedge T})\mid Z_0)/(kq_{\theta}(Z_0)^2)$,
largest where $q_{\theta}$ is small.

\item \emph{VCN.} For reversible dynamics and equilibrium pairs,
$(Z_0,Z_T)$ is exchangeable and the VCN gradient is twice the semi-gradient
of the detached, \emph{unstopped} temporal-difference loss
$\E_\pi[(q_{\bar \theta}(Z_T)-q_{\theta}(Z_0))^2]$. Its minimizer therefore solves $P_T q_{\theta}=q_{\theta}$ on
$\Dom$ rather than $\Pst q_{\theta}=q_{\theta}$, and differs from $q$ through paths that touch
$A\cup B$ before time $T$ and are back in $\Dom$ at time $T$. The bias
vanishes as $T\to0$, where $\mathcal L_{\mathrm{VCN}}/T$ tends to the
Dirichlet form $\E_\pi[|\sigma^\top\nabla q_{\theta}|^2]$ minimized by $q$
\cite{vanden2006towards,khoo2019solving,li2019computing}. Neither property
survives outside this setting: with non-equilibrium pairs (e.g.\ shooting
from previously sampled configurations \cite{tang2026breaking})
exchangeability fails and the minimizer is no longer characterized, and if
the pairs are stopped at $A\cup B$ while both endpoints are differentiated,
$q$ is not even a stationary point \cite[Thm.~E.1]{du2026rare}.
\end{enumerate}
\end{srem}

\subsection{Theoretical guarantees of \name{}}
\label{sec:si-ours}

This subsection derives the \name{} objective, Eq.~\eqref{eq:J_final}, from the squared BKE residual,
Eq.~\eqref{eq:J_initial}, and analyzes it. It\^o's lemma turns the BKE residual integrated along a trajectory into
an expression that involves only $q_\theta$ and $\nabla q_\theta$. The same identity shows that the committor makes
this residual vanish on every trajectory, whatever dynamics generated it, which is what allows off-policy training.
We then relate the objective to those of Section~\ref{sec:si-prior}, explaining why one trajectory per starting
point suffices, and prove that the objective is first-order optimal, with a convergence rate for its gradient flow
(Proposition~\ref{prop:si-V0}).

\paragraph{From the BKE residual to the \name{} objective.} 
Consider the sampling dynamics 
$\dd Z_t=\bt_t\,\dd t+\sqrt{\kappa_t}\,\sigma\,\dd W_t$ introduced in Section~\ref{sec:methods}.

A training segment starts with $Z_0\sim\mu_0$ and ends at $\tau\wedge T$, where $\tau$ is the first
entry time into $A\cup B$ and $T$ the maximum segment length, so it stays in $\Dom=\mathbb{R}^d\setminus(A\cup B)$ up to
time $S$. For a function $q_{\theta}$, It\^o's lemma gives
\begin{align*}
q_{\theta}(Z_{\tau\wedge T})-q_{\theta}(Z_0) &=\int_0^{\tau\wedge T}\Big(\bt_t\cdot\nabla q_{\theta}(Z_t)+\tfrac{\kappa_t}{2}\,\sigma\sigma^\top:\nabla^2 q_{\theta}(Z_t)\Big)\dd t
+\int_0^{\tau\wedge T}\sqrt{\kappa_t}\,\nabla q_{\theta}(Z_t)^\top\sigma\,\dd W_t
\\ &=\int_0^{\tau\wedge T}(\bt_t-\kappa_t b(Z_t))\cdot\nabla q_{\theta}(Z_t)\,\dd t
+\int_0^{\tau\wedge T}\sqrt{\kappa_t}\,\nabla q_{\theta}(Z_t)^\top\sigma\,\dd W_t
+\int_0^{\tau\wedge T}\kappa_t\,\Lop q_{\theta}(Z_t)\,\dd t,
\end{align*}

where the second line adds and subtracts $\kappa_t\,b(Z_t)\cdot\nabla q_{\theta}(Z_t)$ and uses Eq.~\eqref{eq:generator}.
Hence the residual
\begin{equation}
R(q_{\theta})=q_{\theta}(Z_{\tau\wedge T})-q_{\theta}(Z_0)-\int_0^{\tau\wedge T}(\bt_t-\kappa_t b(Z_t))\cdot\nabla q_{\theta}(Z_t)\,\dd t
-\int_0^{\tau\wedge T}\sqrt{\kappa_t}\,\nabla q_{\theta}(Z_t)^\top\sigma\,\dd W_t,
\label{eq:si-residual}
\end{equation}
which involves only $q_{\theta}$ and $\nabla q_{\theta}$, equals the BKE residual integrated along the segment:
\begin{equation}
\;R(q_{\theta})=\int_0^{\tau\wedge T}\kappa_t\,\Lop q_{\theta}(Z_t)\,\dd t\;
\label{eq:si-generator}
\end{equation}
Observe that Eq.~\eqref{eq:J_initial} defines the Pharos objective as $\mathcal J(q_{\theta})=\E[R(q_{\theta})^2]$, and the equalities in \eqref{eq:si-residual} and \eqref{eq:si-generator} immediately prove the equivalence between Eqs.~\eqref{eq:J_initial} and \eqref{eq:J_final}.

Since $\Lop q=0$ in $\Omega$ and the segment stays in $\Omega$ up to time ${\tau\wedge T}$, Eq.~\eqref{eq:si-generator} gives $R(q)=0$ on every segment, whatever the drift, noise multiplier and initial law that generated it. Hence, $\mathcal J(q)=0$ for any training distribution, which is what allows off-policy training: segments generated under earlier controls, intermittent control or other known changes of the dynamics can be used as they are, provided the drift and noise multiplier that generated them are recorded and used in Eq.~\eqref{eq:si-residual}.

\paragraph{Relation to the objectives of Section~\ref{sec:si-prior}.} 
Consider first reference sampling ($\bt_t=b(Z_t)$, $\kappa_t=1$), the setting of objectives (i)--(iii). The drift integral in
Eq.~\eqref{eq:si-residual} then vanishes, and $R(q_{\theta})=q_{\theta}(Z_{\tau\wedge T})-q_{\theta}(Z_0)-M(q_{\theta})$ with
$M(f)=\int_0^{\tau\wedge T}\nabla q_{\theta}(Z_t)^\top\sigma\,\dd W_t$. Since $M(q_{\theta})$ has zero mean given $Z_0$,
$\E[R(q_{\theta})\mid Z_0]=(\Pst q_{\theta}-q_{\theta})(Z_0)$. Splitting $\E[R(q_{\theta})^2]$ into the squared conditional mean and the conditional
variance gives
\begin{equation}
\mathcal J(q_{\theta})=\mathcal J_{\mathrm{SG}}(f)
+\E\big[\operatorname{Var}\big(q_{\theta}(Z_{\tau\wedge T})-M(q_{\theta})\,\big|\,Z_0\big)\big].
\label{eq:si-relation}
\end{equation}
The second term is nonnegative and vanishes at $q_{\theta}=q$, because $q(Z_{\tau\wedge T})-M(q)=q(Z_0)$ pathwise by
Eq.~\eqref{eq:si-generator}. Hence $\mathcal J$ has the same minimizer as the semigroup loss but, unlike it, is an
expectation over single trajectories, so one trajectory per starting point gives an unbiased estimate of the loss
and of its gradient.

The stochastic integral $M(q_{\theta})$ is what makes this work. Without it, the single-trajectory loss
$\E[(q_{\theta}(Z_{\tau\wedge T})-q_{\theta}(Z_0))^2]$ equals $\mathcal J_{\mathrm{SG}}(q_{\theta})+\E[\operatorname{Var}(q_{\theta}(Z_{\tau\wedge T})\mid Z_0)]$, whose second
term depends on $q_{\theta}$ and is positive at $q$, so $q$ is no longer a stationary point
(Remark~\ref{rem:si-prior}(iii)). The previous objectives get around this in different ways: (i) estimates the
conditional mean with two independent trajectories, (ii) detaches the target and averages over a few trajectories, and (iii)
relies on the exchangeability of equilibrium pairs, at the cost of an unstopped fixed point. In $\mathcal J$,
$M(q_{\theta})$ acts instead as a control variate: it leaves $\E[R(q_{\theta})\mid Z_0]$ unchanged and, at $q_{\theta}=q$, cancels the
fluctuations of $q(Z_{\tau\wedge T})$ exactly. For general sampling ($\bt_t\ne b(Z_t)$ or $\kappa_t\ne1$), Eq.~\eqref{eq:si-relation} no longer holds, but the drift
integral in Eq.~\eqref{eq:si-residual} is precisely the correction that keeps $R(q)=0$ pathwise.

\paragraph{Detached target.}

In Eq.~\eqref{eq:J_final}, the terminal value is $q_{\bar\theta}(Z_{\tau\wedge T})$, where $\bar\theta$ is a copy of $\theta$
through which no gradient flows (a stop-gradient), or the boundary value when $Z_{\tau\wedge T}$ lies in $A\cup B$. The loss
value is unchanged, but its gradient omits the contribution of $q_{\bar\theta}(Z_{\tau\wedge T})$. In practice, $\bar\theta$ is an exponential
moving average of $\theta$, again without gradient, so the target lags the trained network and changes more slowly.
Because
$R(q)=0$ pathwise, the committor is a fixed point of this detached update under any frozen sampling law; we do not
prove that it is the only one. The analysis below is for the \emph{undetached} objective, in which $q_{\theta}(Z_{\tau\wedge T})$ is
differentiated, and shows that there $q$ is the unique stationary point.

\begin{sprop}[First-order optimality and convergence of the \name{} objective]
\label{prop:si-V0}
Fix a frozen sampling law with locally bounded drift $\bt$, $\sigma\sigma^\top$ positive definite, segment length
$T\in(0,\infty)$, and $\mu$ with full support on $\Dom$.
\begin{enumerate}[label=(\alph*), leftmargin=*]
\item $\mathcal J$ is a convex quadratic functional with $\mathcal J(q)=0$
and, for every candidate $f$, 
\begin{equation}
\partial_\epsilon\mathcal J\big(f+\epsilon(q-f)\big)\big|_{\epsilon=0}=-2\mathcal J(f).
\label{eq:si-descent}
\end{equation}
\item $q$ is the unique stationary point of $\mathcal J$.
\item Let $\langle\cdot,\cdot\rangle$ be any inner product on candidates
for which the gradient flow $(f_t)_{t\ge0}$ of $\mathcal J$ exists. Then
$\|f_t-q\|$ is nonincreasing and
\begin{equation}
\mathcal J(f_t)\le\frac{\|f_0-q\|^2}{4t}.
\label{eq:si-rate}
\end{equation}
\end{enumerate}
\end{sprop}

\begin{proof}
(a) $R$ is linear with $R(q)=0$, so
$\mathcal J(f+\epsilon(q-f))= (1-\epsilon)^2\mathcal J(f)$, which gives
\eqref{eq:si-descent}; $\mathcal J(f)=\E[R(f)^2]$ is a nonnegative
quadratic form, hence convex.

(b) At a stationary point the derivative in the direction $q-f$ vanishes,
so $\mathcal J(f)=0$ by \eqref{eq:si-descent}, i.e.\
$\int_0^{\tau\wedge T}\kappa_t\,\Lop f(Z_t)\,\dd t=0$ almost surely by \eqref{eq:si-generator}.
Suppose $\Lop f(z_0)\ne0$ for some $z_0\in\Dom$; by continuity,
$|\Lop f|\ge c>0$ with constant sign on a ball $U\subset\Dom$ around $z_0$.
Since $\mu$ has full support and the sampling diffusion is nondegenerate,
with positive probability $Z_0\in U$ and $Z_t\in U$ for all $t\le T$. On
this event ${\tau\wedge T}=T$ and $|\int_0^{\tau\wedge T}\kappa_t\Lop f\,\dd t|\ge\kappa_{\min}cT>0$, a contradiction.
Hence $\Lop f=0$ in $\Dom$. Under the reference dynamics $f(Z_{t\wedge\tau})$
is then a bounded martingale, so
$f(z)=\E_z[f(Z_{t\wedge\tau})]\to\E_z[\mathbf 1_B(Z_\tau)]=q(z)$ as $t\to\infty$, by
dominated convergence.

(c) Let $e_t=f_t-q$. Since $R(f_t)=R(e_t)$,
$\langle\nabla\mathcal J(f_t),e_t\rangle
=\partial_\epsilon\mathcal J(f_t+\epsilon e_t)|_{\epsilon=0}
=2\E[R(f_t)R(e_t)]=2\mathcal J(f_t)$, so
$\frac{\dd}{\dd t}\frac12\|e_t\|^2=-2\mathcal J(f_t)\le0$. Integrating,
$\int_0^t\mathcal J(f_s)\,\dd s\le\frac14\|e_0\|^2$, and since
$\mathcal J(f_t)$ is nonincreasing along the flow,
$t\,\mathcal J(f_t)\le\int_0^t\mathcal J(f_s)\,\dd s$.
\end{proof}

\begin{srem}[Scope of Proposition~\ref{prop:si-V0}]
\label{rem:si-V0}
Part~(a) says that, from any candidate $f$, moving in the direction of the committor, $q-f$, decreases the loss at a
rate equal to twice the loss itself. Part~(c) turns this into a convergence rate: along the gradient flow, the loss
decays at least as $1/t$, for any choice of inner product. The decay becomes exponential if the loss controls the
distance to the committor, i.e.\ if $\E[R(e)^2]\ge c\|e\|^2$ for all $e=f-q$.
The proposition is stated in function space, with exact expectations and exact time integration. Three gaps separate
it from training in practice: \emph{(i)} a neural network can only move within its parameterization, so the direction $q-f_\theta$ may not be available; \emph{(ii)} the replay buffer is finite, so the loss does not determine $f$ outside the sampled region;
\emph{(iii)} time discretization makes $R(q)$ nonzero, so the minimizer of the discrete loss is biased, with a bias that
vanishes as $\Delta t\to0$.
Finally, the proof of (b) assumes that $\sigma\sigma^\top$ is positive definite. For underdamped Langevin dynamics this
fails, because the noise acts only on the velocities. Eq.~\eqref{eq:si-generator} and parts~(a) and~(c) still hold,
but part~(b) would require a hypoelliptic version of the support argument, showing that segments remain near any
phase point with positive probability, which we do not give here.
\end{srem}

\providecommand{\authorcheck}[1]{\par\noindent\textbf{Author check.} \textit{#1}\par}

\section{System setup}
\label{sec:si-systems}

The molecular systems and network inputs are introduced in Methods.
Here we give the simulation parameters and basin definitions in more detail.
All systems use underdamped Langevin dynamics.

\subsection{Alanine dipeptide}
\label{sec:si-sys-adp}

Alanine dipeptide contains 22 atoms and uses CHARMM27 in vacuum.
Nonbonded interactions have no cutoff, and bonds are unconstrained.
The setup uses BAOAB integration at 300\,K, with friction
$0.3$\,ps$^{-1}$ and a 1\,fs step.
The reactant and product basins are periodic boxes in $(\phi,\psi)$.
Their centers are $(-150^\circ,170^\circ)$ and $(90^\circ,-50^\circ)$.
Each angle lies within $10^\circ$ of its basin center.
The committor gives the probability of reaching C7ax before C7eq. Our state definition and system set up follow current existing literature \cite{chen2026welpweightedensemble,mitchell2024committor}.

\subsection{Chignolin}
\label{sec:si-sys-chig}

Chignolin CLN025 contains 166 protein atoms, 2,100 TIP3P waters
and two Na$^+$ ions.
The force field is CHARMM27, and the cubic box has side 4.12009\,nm.
Electrostatics use PME with a 0.95\,nm real-space cutoff.
Bonds to hydrogen are constrained, and water molecules are rigid.
BAOAB dynamics use 340\,K, friction $0.3$\,ps$^{-1}$ and a 2\,fs step.

Following previous studies\cite{chen2026welpweightedensemble,lazzeri2023molecular}, we measure folding with 74 native heavy-atom contacts:
\begin{equation}
Q(x)=\frac{1}{74}\sum_{(i,j)}
\frac{1}{1+\exp[\beta(r_{ij}(x)-1.8r^0_{ij})]},
\qquad \beta=50\,\mathrm{nm}^{-1}.
\end{equation}
Here $r^0_{ij}$ is the distance in the native reference.
The folded basin $A$ has $Q\geq0.99$.
The unfolded basin $B$ has $Q\leq0.01$.
Thus, $q$ is the probability of unfolding first.

\subsection{Trp-cage}
\label{sec:si-sys-tc}

Trp-cage TC5b uses the CHARMM36m system of
Ref.~\citenum{strahan2021long}.
It contains 304 protein atoms, 3,732 CHARMM TIP3P waters,
ten K$^+$ ions and eleven Cl$^-$ ions.
The cubic box has side 4.83853\,nm.
Electrostatics use PME with a 1.2\,nm cutoff.
Lennard-Jones interactions switch off between 1.0 and 1.2\,nm.
Bonds to hydrogen are constrained, and water molecules are rigid.
BAOAB dynamics use 300\,K, friction $0.1$\,ps$^{-1}$ and a 2\,fs step.

The states follow Ref.~\citenum{strahan2021long}.
Let $\rho_\alpha$ and $\rho_3$ be the C$\alpha$ RMSDs of
residues 2-9 and 11-15, respectively, relative to the folded reference.
With all distances in \AA, the unfolded basin is
\begin{equation}
A:\quad
\frac{|\rho_\alpha-2.15|^3}{0.008}
+\frac{|\rho_3-2.8|^3}{0.125}<1.
\end{equation}
The folded basin is
\begin{equation}
B:\quad
\frac{(\rho_\alpha-0.3)^2}{0.0289}
+\frac{(\rho_3-0.3)^2}{0.04}<1,
\qquad d_{\mathrm{C}\alpha_1,\mathrm{C}\alpha_{20}}<17.
\end{equation}
The committor gives the probability of folding first.

\subsection{Host-guest binding}
\label{sec:si-sys-hg}

The OAMe-G2 system contains 211 solute atoms, 2,100 TIP3P waters
and nine Na$^+$ ions.
We use the reference GAFF/RESP parameters and funnel
restraint\cite{rizzi_role_2021}.
Electrostatics use PME with a 1.0\,nm cutoff.
Water molecules are rigid, and the solute is unconstrained.
BAOAB dynamics use 300\,K, friction $1$\,ps$^{-1}$ and a 1\,fs step.
For the guest height $h$ along the host axis,
$A$ is $0<h<0.65$\,nm and $B$ is $h>1.60$\,nm.
The committor gives the probability of dissociation first. These thresholds follow Ref.~\citenum{zhang2026navigatingcommittorlandscapebiomolecules}

\subsection{Ion pairs}
\label{sec:si-sys-ions}

Each system contains one ion pair and 370 rigid TIP3P waters.
Ion parameters are the TIP3P-specific Joung-Cheatham
parameters\cite{joung2008}.
Electrostatics use PME with a 1.0\,nm cutoff.
The Lennard-Jones cutoff is also 1.0\,nm.
OVRVO dynamics use 300\,K, friction $1$\,ps$^{-1}$ and a 2\,fs step.
Starting from the NaCl setup, other salts replace the two ions.
Preparation comprises minimization, 200\,ps of NVT,
1\,ns of NPT at 1\,bar, and 50\,ps of NVT.
The production box uses the mean box length from the second half
of the NPT run.

The basins are $A:r\leq R_A$ and $B:r\geq R_B$,
where $r$ is the ion separation.
The thresholds follow Ref.~\citenum{jung2023machine}.
The committor gives the probability of dissociation first.

\begin{table}[ht]
\centering
\caption{\textbf{Ion-pair basin thresholds.} All distances are in nm.}
\label{tab:ion-pairs}
\label{tab:ion pairs}
\begin{tabular}{lcccccc}
\toprule
Salt  & LiCl & LiI  & NaCl & NaI  & CsCl & CsI  \\
\midrule
$R_A$ & 0.23 & 0.26 & 0.27 & 0.31 & 0.34 & 0.38 \\
$R_B$ & 0.48 & 0.53 & 0.53 & 0.59 & 0.60 & 0.68 \\
\bottomrule
\end{tabular}
\end{table}

Training also includes virtual monovalent ions.
Candidate sizes, well depths and masses are sampled independently
of chemical identity using a Latin hypercube.
The cation ranges are $1/\sigma=2.6$-7.0\,nm$^{-1}$,
$\epsilon=0.05$-2.0\,kJ\,mol$^{-1}$ and $m=7$-133\,Da.
The anion ranges are $1/\sigma=1.7$-2.8\,nm$^{-1}$,
$\epsilon=0.05$-1.0\,kJ\,mol$^{-1}$ and $m=30$-130\,Da.
Well depths and masses are sampled on logarithmic scales.
Six virtual cations are paired with Cl$^-$ and I$^-$.
Three virtual anions are paired with one virtual cation.
This gives 15 candidate salts.

For virtual salts, $R_A=0.75\sigma_{\rm pair}$ and
$R_B=1.55\sigma_{\rm pair}$.
Here $\sigma_{\rm pair}$ is the mean of the two ion diameters.
A salt is retained if contact occupancy is at least 5\%
in at least one of two unbiased 20\,ps relaxation runs.
An unbiased reference configuration in $B$ is also required.
Seven salts remain: C1A0, C1A2, C1Cl, C4Cl, C4I, C5Cl and C5I.

\begin{table}[ht]\centering
\caption{\textbf{Virtual ions in the retained salts.}
C and A denote virtual cations and anions, with charges $+1$ and $-1$.}
\label{tab:ion-virtual}
\begin{tabular}{lccc}\toprule
Ion & $\sigma$ (nm) & $\epsilon$ (kJ\,mol$^{-1}$) & Mass (Da)\\\midrule
C1 & 0.27027 & 0.23248 & 104.06\\
C4 & 0.16949 & 0.06800 & 8.95\\
C5 & 0.15083 & 0.12569 & 39.00\\
A0 & 0.53097 & 0.08243 & 38.31\\
A2 & 0.38217 & 0.22361 & 62.45\\
\bottomrule\end{tabular}
\end{table}

The full model uses all six physical salts.
Five further models exclude every salt containing one selected ion.
This exclusion also applies to virtual salts containing a physical anion.
All models use the same training settings.

\subsection{Hydrogen cyanide}
\label{sec:si-sys-hcn}

The liquid HCN system contains 64 HCN units in a cubic box of side 1.5426\,nm.
The aqueous system contains one \ce{CN-}, one HCN and one Na$^+$
in 64 waters, in a cubic box of side 1.2242\,nm.
Both systems use UMA-s-1.2.1 with the OMol
task\cite{woodUMAFamilyUniversal2025,levineOpenMolecules20252026}.
Langevin dynamics use 300\,K, friction $10$\,ps$^{-1}$ and a 0.5\,fs step.

The product basin contains any C--C bond of at most 1.60\,\AA.
In liquid HCN, the reactant basin contains the neutral liquid
or a free \ce{CN-}/\ce{HCNH+} pair.
A cyanide is free when its carbon is at least 2.80\,\AA\
from every other carbon.
In water, the reactant basin has a C--C distance of at least 3.00\,\AA.
The C--N-linked isomer belongs to the reactant basin.
The committor gives the probability of reaching the C--C-bonded product first.

Molecular identities are assigned from the bond graph.
C--C bonds use a cutoff of 1.75\,\AA.
Other heavy-atom bonds use 1.60\,\AA.
Each hydrogen is assigned to its nearest heavy atom within 1.30\,\AA.

\section{Networks, training and evaluation}
\label{sec:si-exp}

\subsection{Networks}
\label{sec:si-networks}

All networks learn the phase-space committor $q(x,v)$.
System compositions, dynamics and basin definitions are given in
Supplementary Section~\ref{sec:si-systems}.

\paragraph{Alanine dipeptide and chignolin.}
Both models use a multilayer perceptron (MLP).
Positions enter through heavy-atom pair distances.
Velocities enter through the rates of change of these distances:
\begin{equation}
d_{ij}=|x_i-x_j|,\qquad
\dot d_{ij}=\frac{(x_i-x_j)\cdot(v_i-v_j)}{d_{ij}}.
\end{equation}
The distances and their rates are concatenated in a fixed atom-pair order
and passed to the MLP.

Alanine dipeptide uses all 45 pairs among its ten heavy atoms.
Its five hidden layers contain 64, 128, 256, 128 and 64 units.
Chignolin uses all 4,278 pairs among its 93 heavy atoms.
Its four hidden layers contain 256, 256, 128 and 64 units.
Pair distances are expressed in nm and their rates of change in nm\,ps$^{-1}$.

Both MLPs use SiLU activations and output two positive scalars,
which are combined into $q_\theta$ as defined below.

\paragraph{Trp-cage.}
We use a  PaiNN network\cite{painn2021} taking positions, velocities
and residue identities of the 20 C$\alpha$ atoms.
These atoms are the graph nodes.
Each velocity is divided by the atom's thermal speed,
$\sqrt{k_{\mathrm B}T/m_i}$.
The network has three interaction layers with 96 scalar
and 96 vector features per node.
Distances are encoded by 20 Gaussian functions,
with a smooth cutoff at 1.0\,nm.

To form the output, we concatenate each node's scalar features
with the magnitudes of its vector features.
A shared MLP maps these 192 values to 96 values per node.
We average these outputs over the 20 C$\alpha$ atoms.
An output MLP with two hidden layers of 96 units maps this average
to two positive scalars, which enter the final expression for $q_\theta$.

\paragraph{Ion pairs.}
We use a  PaiNN network\cite{painn2021} to handle both ions and all water atoms.
Each velocity is divided by the atom's thermal speed,
$\sqrt{k_{\mathrm B}T/m_i}$.
Each ion enters through its physical properties rather than
a chemical-species label.
Its input is $(Q_i/\sigma_i,\sigma_i,\epsilon_i,m_i)$,
where $Q_i$ is its charge, $\sigma_i$ its Lennard-Jones size,
$\epsilon_i$ its well depth, and $m_i$ its mass.
Water oxygens and hydrogens have distinct atom labels.

The network has three interaction layers with 32 scalar
and 32 vector features per atom.
The network use a cutoff of 0.6\,nm.
The message passing includes pair distances,
relative velocities, and Coulomb and Lennard-Jones energies.
Two sets of 20 Gaussian functions encode the distances.
One uses distances in nanometers.
The other uses distances divided by the ion pair's Lennard-Jones
size, $(\sigma_++\sigma_-)/2$ for normalization.

The final output uses the learned features of the two ions.
Water contributes to these features through message passing;
its final features are not included directly in the output layer.
For each ion, we retain its scalar features and convert
each vector feature into two scalars: its magnitude and
its projection onto the axis joining the ions.
We concatenate these quantities with the ion separation,
relative radial velocity, and the pair's Lennard-Jones
size and well depth.
An output MLP with two hidden layers of 32 units and SiLU
activations maps this input to one scalar.
A sigmoid converts it to $q_\theta$.
The prediction is set to zero in basin $A$ and one in basin $B$.

\paragraph{Host-guest binding.}
We use a PaiNN network\cite{painn2021} taking the positions
and velocities of the 143 solute heavy atoms, selected water
oxygens and all nine sodium ions.
We translate and rotate each configuration to align the host
with a fixed reference.
Velocities undergo the same rotation.
Each velocity is divided by the atom's thermal speed,
$\sqrt{k_{\mathrm B}T/m_i}$.
Element labels distinguish the atoms, with separate labels
for water and solute oxygens.

We include the 96 water oxygens closest to the 73 host atoms
lining the cavity.
For each water oxygen or sodium ion, $d$ is its distance
to the nearest lining atom.
We weight messages to and from these atoms by $w(d)$.
The weight is one for $d\leq0.40$\,nm and zero for
$d\geq0.45$\,nm.
Between these distances,
\begin{equation}
w(d)=\frac12\left[
1+\cos\left(
\pi\frac{d-0.40\,\mathrm{nm}}{0.05\,\mathrm{nm}}
\right)\right].
\end{equation}
This gradually removes an atom's contribution as it moves
away from the cavity.
The network has three interaction layers with 32 scalar
and 32 vector features per atom.
Distances are encoded by 20 Gaussian functions,
with a smooth cutoff at 0.6\,nm.
Messages also include the relative velocities of
neighboring atoms.

To form the output, we concatenate the scalar features
and vector magnitudes of the 143 solute heavy atoms
in a fixed atom order.
Water and sodium contribute to these features through
message passing; their final features are not included
directly in the output layer.
An output MLP with hidden layers of 64 and 32 units
and SiLU activations maps the combined features to two scalars.
These are combined into $q_\theta$ as defined below.

\paragraph{Hydrogen cyanide.}

We use a PaiNN network\cite{painn2021} taking the positions and velocities
of all atoms in the box. 
Each atom is identified by its element. 
Each velocity is divided by the atom's thermal speed,
$\sqrt{k_{\mathrm B}T/m_i}$. 
The network uses a cutoff of 0.5\,nm. Each layer maintains 32 scalar and 32 vector features per atom.
20 sine-based radial functions represent interatomic distances.

To form the output, we use the learned scalar atomic features of each C--C pair in the
entire system.
We input the sum and element-wise product of the features of the two carbons, together with 16 radial functions
of their interatomic distance, to an output MLP.
This MLP with one hidden layer of 32 units and produces
a value $u_{ij}$ for each C--C pair. 
We also average the scalar features over all atoms and use a second MLP, with one hidden layer of 32 units, to map this average to a value $u_0$.
Finally, we combine $u_0$ and all carbon-pair values
$u_{ij}$ to obtain the committor:
\begin{equation}
S=e^{u_0}
+\sum_{i<j;\,i,j\in{\rm C}}f_c(d_{ij})e^{u_{ij}},
\qquad
q_\theta=\frac{S}{1+S}.
\end{equation}
Here $d_{ij}$ is the C--C distance, while $f_\text c$ is a cutoff function that decreases smoothly to zero at 0.45\,nm in liquid HCN
and 0.60\,nm in the aqueous solution.

\paragraph{Output and basin information.}
For alanine dipeptide, chignolin, Trp-cage and host-guest binding,
the two scalar outputs are made positive using
$\operatorname{softplus}+10^{-6}$.
Denoting these positive values by $a_\theta$ and $b_\theta$,
we calculate
\begin{equation}
q_\theta=
\frac{w_Aa_\theta}{w_Aa_\theta+w_Bb_\theta}.
\end{equation}
The weights incorporate the following structural information.
Alanine dipeptide uses the periodic distances in $(\phi,\psi)$
to the two basin centers as $w_A$ and $w_B$.
For both proteins, the output weights depend on the fraction
of native contacts.
Each native atom pair $(i,j)$ contributes
\begin{equation}
c_{ij}(x)=
\frac{1}{1+\exp\!\left[\beta\left(d_{ij}(x)-\lambda d_{ij}^{0}\right)\right]},
\qquad \beta=50\,\mathrm{nm}^{-1}.
\end{equation}
Here $d_{ij}(x)$ is the current pair distance and $d_{ij}^{0}$
is its distance in the native reference.
The contribution decreases continuously as the atoms separate.
The contact fraction is the average of $c_{ij}$ over the
specified native pairs.
For the output weights, distances are evaluated using
unwrapped protein coordinates.
Chignolin uses a fixed set of 74 native heavy-atom pairs
with $\lambda=1.8$.
Denoting their contact fraction by $Q$, we set
$(w_A,w_B)=(1-Q,Q)$. 
For Trp-cage, native pairs are C$\alpha$ atoms separated
by at least three positions in the sequence and less than
0.8\,nm in the folded reference.
We use $\lambda=1.2$ and denote the resulting contact
fraction by $Q_{\rm CA}$.
The output weights are
$(w_A,w_B)=(Q_{\rm CA},1-Q_{\rm CA})$.
The weight order differs because $q$ describes unfolding
for chignolin and folding for Trp-cage.
For numerical stability, each weight is bounded below
by $10^{-6}$.
Host-guest binding uses $w_A=w_B=1$.
For these four systems, training and control use
$q_\xi=\xi+(1-2\xi)q_\theta$
with $\xi=e^{-5}$ following Ref.~\citenum{du2026rare} unless specified below.
For the ion system, to reduce prior knowledge and to let the network learn transferability on its own, we do not use this preconditioning and use only a sigmoid to convert one scalar output to $q_\theta$.
All reported evaluations use the unregularized committor $q_\theta$ instead of $q_\xi$.

\subsection{Training}
\label{sec:si-exp-common}
\label{sec:si-exp-systems}

\paragraph{Common procedure.}
Training alternates between controlled simulation and network updates.
Stored trajectories are reused in a replay buffer.
The loss is evaluated over short trajectory intervals, which we will refer to as segments.
Known basin outcomes replace predictions at basin entry.
We use a learning rate
of $10^{-3}$ and clip the gradient norm at 10.
We use an EMA with rate 0.99.

Our training loss follows Eq.~\ref{eq:J_final}.
Discretizing Eq.~\ref{eq:si-residual} gives
$R_\theta(s,e)\simeq A+B_1+B_2+C$, where
\begin{align}
A &= q_{\bar\theta}(x_e,v_e)-q_\theta(x_s,v_s),
\nonumber\\
B_1 &= -\sum_{k=s}^{e-1}
\bigl(\widetilde b_k-b(x_k,v_k)\bigr)
\cdot\nabla_{x,v}q_\theta(x_k,v_k)\,\Delta t_k,
\nonumber\\
B_2 &= \sum_{k=s}^{e-1}
(\kappa_k-1)b(x_k,v_k)
\cdot\nabla_{x,v}q_\theta(x_k,v_k)\,\Delta t_k,
\nonumber\\
C &= -\sum_{k=s}^{e-1}
\sqrt{\kappa_k}\,
\nabla_{x,v}q_\theta(x_k,v_k)^\top\sigma\,\Delta W_k.
\label{eq:si-residual-discrete}
\end{align}
Here $\nabla_{x,v}$ denotes the gradient with respect
to positions and velocities.
The time between consecutive stored states is $\Delta t_k$,
and $\Delta W_k$ is the corresponding Brownian increment.
The drift $b$ describes the reference dynamics,
while $\widetilde b$ describes the sampling dynamics.
The factor $\kappa_k$ multiplies the noise variance.
The term $A$ is the endpoint difference.
The drift correction is split into $B_1$, which accounts
for control and force modifications, and $B_2$, which accounts
for the change in noise variance.
When $\kappa_k=1$, $B_2$ vanishes.
The term $C$ is the stochastic contribution.

Both drift contributions can also be approximated using
finite changes in the network prediction.
For a drift $a=(a_x,a_v)$, we use
\begin{equation}
a\cdot\nabla_{x,v}q(x,v)\,\Delta t
\simeq
q(x+a_x\Delta t,v+a_v\Delta t)-q(x,v).
\label{eq:si-drift-finite}
\end{equation}
For $B_1$, the drift is $a=\widetilde b-b$;
for $B_2$, it is $a=b$.
The corresponding factors are $-1$ and $\kappa-1$,
respectively.

\paragraph{Alanine dipeptide.}

\textit{Initialization and restarts.}
We initialize trajectories from the two reference conformers
for basins $A$ and $B$.
We add Gaussian perturbations with standard
deviation 0.001\,nm to positions and draw thermal velocities at 300\,K.
These starts undergo up to 200\,fs of dynamics,
stopping once they lie outside both basins.

During training, 70\% of starts follow this procedure.
The remaining 30\% reuse stored positions and velocities
near predicted $q=0.5$.

\textit{Simulation.}
Each round starts 256 trajectories.
At the start of each  {trajectory}, we select basin $B$
as the control target with probability proportional to our learned $q_\xi(x,v)$,
and basin $A$ otherwise.
The target remains fixed within the  {trajectory}.
The current network computes the corresponding control
at every MD step.
Each  {trajectory} lasts up to 250\,fs and stops upon entering
either basin.
In half of the simulations, we multiply the physical forces
by 0.5 to promote transitions between the basins
(Fig.~\ref{fig:overview}).
The thermostat noise remains unchanged.
If a trajectory  {reaches} neither  {basin,} {we} start the next  {trajectory} from its final position and velocities within the same round.  We allow up to four additional continued simulations.  At the start of each new trajectory, we {sample} the control target again using the same rule from our learned committor. 
After this sampling stage, we store trajectories in the buffer.

\textit{Buffer.}
The buffer stores up to 64,000  {trajectories}.
When full, it discards the oldest {trajectories} to make room
for new ones (FIFO).
For each continuation, we also combine the end of the
preceding  {trajectory} with the beginning of the new  {trajectory}.
This additional stored  {trajectory} allows loss  {segments}
to cross the original {trajectory} boundary.

\textit{Segments used in loss.}
After collecting trajectories, we perform 400 network updates.
For each update, we randomly sample 256  {trajectories} from the buffer,
with equal probability for each stored  {trajectory}.
We divide each  {trajectory} into consecutive 50\,fs  {segments}
and calculate the loss on each  {segment}.
If the trajectory enters a basin, the current {segment}
ends at that point.
We average the losses over all these {segments}
to update the network.
Finally, for  {segments} ending at basin entry,
we use its true committor value to replace the predicted endpoint value in the loss.

\paragraph{Chignolin and Trp-cage.}

\textit{Initialization and restarts.}
For chignolin, we run preparation simulations from
the folded and unfolded reference conformers at 340\,K.
We also run simulations at 510\,K from the folded conformer
to promote unfolding.
For Trp-cage, the unfolded basin contains diverse conformations.
We therefore generate unfolded starting configurations
through preparation simulations from the folded reference.
We run simulations at 300 and 450\,K.
We also run simulations at 300\,K with protein-protein
electrostatic and attractive Lennard-Jones interactions
multiplied by 0.4 to promote unfolding.
For both proteins, we retain configurations outside both
basins as training starting points.

During training, 10\% of starts use these prepared configurations.
 45\% use randomly selected stored configurations.
The remaining 45\% use stored configurations selected
to cover predicted committors from 0.05 to 0.95.
All starts draw fresh thermal velocities at 340\,K
for chignolin and 300\,K for Trp-cage.

\textit{Simulation.}
At each training iteration, we simulate 24 trajectories, store the resulting  {trajectories} in the buffer, and update the network using segments {selected} from the buffer trajectories.
If a trajectory reaches no basin, we {start} the {next} trajectory from its final positions and velocities in the next training iteration.  {We}  {continue}  {until}  {either} basin  {is}  {reached} or ten  {consecutive}  {trajectories}  {have}  {been}  {generated.}  {We}  {then}  {choose}  {a}  {new}  {starting}  {state.}

At the start of each {trajectory}, we select basin $B$
as the control target with probability given by
our learned $q_\xi(x,v)$, and basin $A$ otherwise.
The target remains fixed within the {trajectory}.
To reduce communication overhead between the MD engine
and PyTorch, we apply control every 30\,fs for chignolin
and every 300\,fs for Trp-cage.
We multiply the control by ten to strengthen the bias
toward the selected basin.

To promote transitions, we scale protein-protein
electrostatic interactions and the attractive $r^{-6}$
Lennard-Jones term.
We leave these interactions unchanged in 30\% of trajectories.
In the remaining 70\%, we either weaken or strengthen them.
The scaling factors are 0.5 and 1.3 for Trp-cage,
and 0.5 and 2.0 for chignolin.
We select these factors through brief preliminary tests,
checking whether trajectories reach both basins regularly.
For Trp-cage, 35\% of trajectories use each modified setting (weaken or strengthen).
For chignolin, since  basins A and B are very thin following our definition, we further adjust the relative allocation every ten
training iterations, using basin outcomes from the preceding
30 iterations.
The repulsive $r^{-12}$ term, solvent interactions
and thermostat temperature remain unchanged.

Each chignolin  {trajectory} lasts up to 20\,ps with unchanged
forces and 8\,ps with modified forces.
For Trp-cage, these durations are 120 and 8\,ps.
We record states every 30\,fs with unchanged forces
and every 2\,fs with modified forces.
Each  {trajectory} stops upon entering either basin.

\textit{Buffer.}
The buffer stores the latest 1,000 {trajectories} for chignolin
and 2,000 for Trp-cage.
When full, it discards the oldest  {trajectories} to make room
for new ones (FIFO).
For each continuation, we also combine the end of the
preceding  {trajectory} with the beginning of the new  {trajectory}.
This additional stored  {trajectory} allows loss  {segments}
to cross the original  {trajectory} boundary.

When a  {simulation} reaches a basin, we retain its available
 {trajectories} separately, so they remain available after removal
from the FIFO buffer. 
This is because for these larger systems, the commitment is rare. We do not want to lose their signal easily.
We retain up to 2,000 such  {trajectories}.

\textit{{Segments} used in loss.}
After collecting trajectories in each training iteration,
we perform up to 400 network updates for chignolin
and 32 for Trp-cage.
For each update, we sample 32 {trajectories} from the buffer.
One quarter are selected from {trajectories} ending in a basin,
with equal numbers ending in $A$ and $B$ when both are available.
For these {trajectories}, we use their final 512 recorded steps,
or fewer if the  {trajectory} is shorter.
The remaining  {trajectories} are sampled at random,
and we use up to 512 consecutive recorded steps from each.
We divide the selected data into consecutive loss  {segments}.
 {Segments} last 60\,fs with unchanged dynamics' simulation
and 4\,fs with modified forces since their transition is faster.
If the selected data are shorter, we use their full duration.
We use the discretization in Eq.~\ref{eq:si-drift-finite} to calculate $B_1$ in the training loss for both Chignolin and Trp-cage.
We also set $\xi=e^{-12}$ to reduce bias.

\textit{Additional boundary loss.}
For  {segments} ending at basin entry,
we use the its true committor value to replaces the predicted endpoint value in the loss.
We add a squared-error penalty between the unregularized
prediction $q_\theta$ and the known basin labels, 0 or 1.
(equivalently, enforce
$q_\xi=\xi$ in $A$ and $q_\xi=1-\xi$ in $B$).
Each update uses basin-entry configurations from the sampled
segments and eight additional configurations from each basin.
The additional configurations receive fresh thermal velocities.

\paragraph{Ion pairs.}

\textit{Initialization and restarts.}
We train one model on all six physical salts and five models
that each exclude one ion species.
We initialize trajectories from the reference conformers
for basins $A$ and $B$ of each salt.
We add Gaussian perturbations with standard deviation
0.01\,nm to positions and draw thermal velocities at 300\,K.
These starts undergo up to 400\,fs of dynamics,
stopping once they lie outside both basins.

During training, about 40\% of starts follow this procedure.
Another 30\% reuse stored positions and velocities
near predicted $q=0.5$.
Also, as training goes, some trajectories complete transitions
from $A$ to $B$ or from $B$ to $A$.
We sample the remaining 30\% of starts from configurations
along these trajectories, outside both basins,
and draw fresh thermal velocities.
If too few suitable stored configurations are available,
we use the initialization procedure described above.

\textit{Simulation.}
At each iteration of training, we simulate 160 trajectories.
Half of them simulate physical salts
and half simulate virtual salts.
We distribute trajectories approximately evenly among physical salts
and among virtual salts.
At the start of each  {trajectory}, we select basin $B$
as the control target with probability given by
our learned $q_\theta(x,v)$, and basin $A$ otherwise.
The target remains fixed within the  {trajectory}.
We apply control and check whether a basin is reached every 20\,fs.
We record states every 10\,fs.
Each  {trajectory} lasts up to 0.5\,ps
and stops when a basin is reached.

In half of the simulations, we multiply the physical forces
by 0.5 to promote transitions between the basins.
The thermostat noise remains unchanged.
If a trajectory reaches neither basin,  {we}  {start}  {the}  {next}  {trajectory} {from} {its} {final}  {positions} {and}  {velocities} within the same  {round.} 
We allow up to four additional {trajectories.}  {At}  the {start}  {of} {each} {new}  {trajectory,}  {we}  {sample} the control target again using our learned committor.  {The}  {consecutive}  {trajectories}  {last}   at most {2.5\,ps}  {in}  {total.}
We store the resulting  {trajectories} in the buffer
before updating the network.

\textit{Buffer.}
The buffer stores up to 8,000  {trajectories},
whether or not they reach a basin.
When full, it discards the oldest {trajectories} to make room
for new ones (FIFO).
For each continuation, we also combine the end of the
preceding {trajectory} with the beginning of the new  {trajectory}.
This additional stored trajectory allows loss  {segments}
to cross the original {trajectory} boundary.

\textit{{Segments} used in loss.}
After collecting trajectories, we perform 128 network updates.
For each update, we sample 42  {trajectories} from the buffer.
Half come from physical salts
and half from virtual salts.
We distribute these samples approximately evenly among
physical salts and among virtual salts.

For each salt, we sample  {trajectories} with replacement,
with probabilities proportional to
$\max(q_{\max}-q_{\min},10^{-3})$.
Here $q_{\max}$ and $q_{\min}$ are the largest and smallest
predicted committors in the  {segment} last used to calculate
that  {trajectory}'s loss.
This allows more training over the region where $q$ changes more.

From each sampled  {trajectory}, we select one  {segment}
of up to 30\,fs to calculate the loss.
If the  {trajectory} ends in a basin, we choose its final  {segment}
with probability 0.5; otherwise, we choose the starting
point uniformly.
For  {trajectories} that do not end in a basin,
we always choose uniformly.
If a  {trajectory} is shorter than 30\,fs, we use its full duration.
The model directly sets $q=0$ in $A$ and $q=1$ in $B$.
If a loss  {segment} ends at basin entry,
the corresponding value supplies its endpoint label.
We average the losses over the selected  {segments} and we use the discretization in Eq.~\ref{eq:si-drift-finite} to calculate $B_1$ in the training loss.

\paragraph{Host-guest binding.}

\textit{Initialization and restarts.}
We first run simulations from configurations inside
basins $A$ and $B$.
We save configurations after these simulations leave
the basins and use them to initialize training trajectories.
We choose between the two basins with equal probability.
We add Gaussian perturbations with standard deviation
0.001\,nm to solute positions and draw thermal velocities
at 300\,K.
Starts inside either basin undergo up to 400\,fs
of unbiased dynamics, stopping once they lie outside both basins.
Starts that remain inside a basin are discarded.

During training, approximately half of starts follow
this procedure.
The remaining starts use stored positions and velocities
from training trajectories or separately prepared configurations.
We select configurations outside both basins.
We divide the predicted committor range from 0 to 1
into ten equal intervals and sample approximately evenly
from intervals containing available configurations.
If too few suitable configurations are available,
we use the initialization procedure above.

\textit{Simulation.}
At each training iteration, we simulate 64 trajectories,
store the resulting trajectories in the buffer,
and update the network using segments selected
from the stored trajectories.
Each trajectory lasts up to 120\,ps and stops upon
entering either basin.
If neither basin is reached, we start the next trajectory
from its final positions and velocities in the next
training iteration.
We continue until either basin is reached or
25 consecutive trajectories have been generated.
We then choose a new starting state.

At the start of each trajectory, we select basin $B$
as the control target with probability given by
our learned $q_\xi(x,v)$, and basin $A$ otherwise.
The target remains fixed within the trajectory.
We apply control every 300\,fs and multiply its strength
by ten.
At each control step, we cap the magnitude of the
velocity change $\delta v_i$ at $2\sqrt{k_{\mathrm B}T/m_i}$.
Here $m_i$ is the atomic mass and $T=300$\,K.
For water, we then adjust these increments so that each
molecule moves and rotates without changing its internal geometry.

To promote transitions, we run 70\% of new simulations
at 450\,K and the remaining 30\% at 300\,K.
Only the thermostat temperature changes;
physical forces and friction remain unchanged.
If a simulation continues in the next training iteration,
its temperature stays the same.
The loss accounts for the temperature change so that
we still learn the committor at 300\,K.
We record states every 60\,fs.

\textit{Buffer.}
The buffer stores up to 2,400 trajectories.
For each continuation, we also combine the end of the
preceding trajectory with the beginning of the new trajectory.
This additional stored trajectory allows loss segments
to cross the original trajectory boundary.
When replacing old trajectories, we retain at least
300 trajectories ending in each basin whenever that many
are available.

\textit{Segments used in loss.}
After collecting trajectories in each training iteration,
we perform 160 network updates.
For each update, we sample up to 32 trajectories from the buffer.
From each trajectory, we select up to 256 consecutive recorded
steps (15.36\,ps) to calculate the loss.
If a trajectory is shorter, we use the whole trajectory.

One quarter of the sampled trajectories end in a basin,
with equal numbers ending in $A$ and $B$ when both are available.
For these trajectories, we select the final 256 recorded steps
so that the selected data include the known basin outcome.
The remaining trajectories are sampled uniformly from the buffer.
If a selected trajectory also ends in a basin,
we select its final 256 recorded steps with probability 0.5.
Otherwise, we choose the starting point uniformly
among positions where 256 consecutive recorded steps are available.

The buffer may contain several consecutive trajectories
from the same simulation.
We additionally sample up to eight simulations at 300\,K
with at least two trajectories still stored.
We favor simulations ending in a basin,
with balanced sampling of $A$ and $B$ when both are available.
For each simulation, we join its consecutive trajectories
and use the final 1,024 recorded steps (61.44\,ps),
or fewer if the joined data are shorter.

Within the selected data, we calculate the loss over segments
of 60, 120 and 240\,fs, continuing to double the length
while it fits within the selected duration.
For each length, the segments do not overlap.
We place them consecutively backwards from the end
of the selected trajectory.
We also include the full selected duration
unless it is already included.
Each segment receives a weight equal to its duration
divided by the duration of the selected data from which it is taken.
We take a weighted average of the losses over all segments
in the update.

For segments lasting up to 32 recorded steps (1.92\,ps),
we calculate the residual as $A+B+C$.
For longer segments, we omit $C$ and use $A+B$.

For simulations at 450\,K, the thermostat noise variance
is $\kappa=1.5$ times its value at 300\,K and hence the $B_2$ term does not vanish in our loss. We use Eq.~\ref{eq:si-drift-finite} to calculate $B_2$.

 Finally, for segments ending at basin entry, we use its true committor value to replace the predicted endpoint value in the loss.

\paragraph{Hydrogen cyanide.}

\textit{Initialization and data preparation.}
We first generate configurations at different stages of the reaction.
These provide starting points for subsequent simulations.
For liquid HCN, we start from a configuration containing the C--C-bonded product.
We fix this C-C distance and run 10\,ps of NVT dynamics at 300\,K to allow the surrounding molecules to relax.
We then release the distance constraint and use steered MD to pull the two carbons apart over 6\,ps.
A harmonic restraint gradually increases the target C--C distance to 0.35\,nm.
For the aqueous system, we start from HCN and \ce{CN-} in water.
We first run 10\,ps of NVT dynamics at 300\,K,
without constraining the C-C distance.
We then use steered MD to bring the two carbons together over 6\,ps.
The harmonic restraint gradually decreases the target C--C distance to 0.15\,nm.
Both systems use a restraint force constant of
$2\times10^5$\,kJ\,mol$^{-1}$\,nm$^{-2}$.

We repeat this preparation procedure to generate multiple trajectories, each providing a set of starting configurations.
For liquid HCN, we assign 40 preparation trajectories to training, 8 to validation and 16 to testing.
For the aqueous system, these numbers are 32, 8 and 8.

\textit{Simulation.}
From each training and validation run, we pick four starting configurations randomly from each of 6 C–C distance intervals. 
Each starting configuration initializes eight unbiased trajectories with fresh thermal velocities at 300\,K.
Each trajectory stops when it reaches either basin, or after at most 5\,ps.
The network is first trained on these trajectories, for 8,192 updates in liquid HCN and 10,240 updates in water.

In this stage, training continues to 20,480 updates in liquid HCN and 24,576 in aqueous solution. We use four simulation processes to generate controlled trajectories in parallel.
Each process works in rounds and runs independently of the network updates.
The trainer saves the network every 512 updates,
and each round uses the most recently saved network.

At the start of each round, the simulation process uses the network to evaluate two sets of candidate configurations
and picks 16 starting points among them.
The first set contains 128 configurations drawn from a fixed pool,
with Gaussian noise of 0.001\,nm on the positions and fresh thermal velocities at 300\,K.
The pool includes the starting configurations from the preparation runs and additional simulations started from these runs, 
including an umbrella-sampling run along the C--C distance and plain MD in liquid HCN, 
and OPES along the C--C distance and plain MD in aqueous solution.
The second set contains up to 128 configurations with their velocities randomly sampled from earlier controlled trajectories of the simulation process. 
The 16 starting points are drawn uniformly from bins of the predicted committor to ensure that every stage of the reaction is represented.

In each round of controlled sampling, the trajectory is steered toward a control target without tempering on the forces. At the start of each trajectory, we draw its control target at random: 
the product with probability $q_\xi(x,v)$ of the starting configuration and velocity, 
and the reactant with probability $1-q_\xi(x,v)$. 
With this random draw alone, these starts would almost always be steered back into the nearby reactant basin,
and the buffer would contain few paths that leave the reactant side and reach the product. To supplement such trajectories that the rate calculations rely on, for starts with $q_\xi<0.1$, we set the product as the target.
Each trajectory stops when it enters a basin and lasts at most 512\,fs.
Trajectories that reach no basin are discarded.

\textit{Buffer.}
The training buffer and stores the complete paths from which the loss is computed.
It initially holds the uncontrolled trajectories of the training runs and adds trajectories from all four processes every 32 updates of the network.
It keeps at most twice as many paths as the number of the uncontrolled trajectories. When it is full, the oldest paths are removed first (FIFO), so controlled trajectories gradually replace the uncontrolled trajectories.

\textit{{Segments} used in loss.}
Each update of the network samples 16 trajectories uniformly from the buffer with replacement.
For each trajectory, we evaluate the residual with the drift term of Eq.~\eqref{eq:si-drift-finite}
on the whole trajectory and on the trajectory segments starting one quarter, one half and three quarters
of the way along it; all four end at the first basin entry.
The segment losses are averaged within each trajectory and then over the 16 trajectories. The output regularization starts at $\xi=e^{-5}$ and is reduced to $e^{-9}$ and $e^{-13}$
at updates 10,240 and 14,336 in liquid HCN, and 12,288 and 16,384 in aqueous solution.
We use Adam without weight decay, with a learning rate of $10^{-3}$ in liquid HCN
and $10^{-4}$ in aqueous solution.
 
\textit{Additional boundary loss.}
We also add a squared-error penalty that pushes the network output toward the known basin values,
$q_\theta=1$ in the product basin and $q_\theta=0$ in the reactant basin.
In each update, the penalty is evaluated at four trajectory endpoints, two in the product basin and two in the reactant basin, drawn from the endpoints of all trajectories collected so fars. This keeps the predicted committor correct deeper inside the basins.

\textit{Additional settings.}
In aqueous solution, the predicted committor on the reactant side spans many orders of magnitude,
down to about $10^{-10}$. The residual above measures absolute errors in $q$,
so it hardly refines such small values, and after the main training
$\operatorname{logit}q$ was nearly flat for C-C distances between 2.1 and 3.0\,\AA.
A flat $\operatorname{logit}q$ cannot order these configurations, which matters
when $s=\operatorname{logit}q$ is used for umbrella sampling.
Therefore, during the final 4,096 updates in aqueous solution, we add a logarithm residual, equivalent to the value matching objective\cite{du2026rare}. 

We evaluate it on the same trajectory segments as the main residual,
average its square within each trajectory and then over the sampled trajectories,
and add it to the loss with weight 0.003.
Liquid HCN does not use this term.

\subsection{Evaluation}
\label{sec:si-eval}

We evaluate the unregularized model output $q_\theta(x,v)$.
The comparisons below use the configuration committor
$q_\theta(x)$, obtained by averaging over velocities.
For each test configuration $x$, we draw independent velocity samples from the Maxwell-Boltzmann distribution at the system temperature.
The velocities satisfy the molecular constraints
where applicable.
We calculate
\begin{equation}
q_\theta(x)\approx
\frac{1}{M}\sum_{m=1}^{M}q_\theta(x,v^{(m)}).
\end{equation}

 \paragraph{Alanine dipeptide.} \textit{Test configurations, shooting and baselines.}
The test set contains 54,520 configurations,
with shooting committors distributed approximately uniformly between 0 and 1.
We use $M=64$ velocity samples per configuration. We compare predictions from \name{} and the semigroup method~\cite{mitchell2024committor} against the same shooting test set in literature \cite{chen2026welpweightedensemble}.The test set contains ~360,000 ADP conformations and each conformation was simulated 1000 times to record is empirical committot value.  

\paragraph{Chignolin.}\textit{Test configurations, shooting and baselines.}
The test set contains 2,134 configurations.
The Chignolin (CLN025) test set is constructed from equilibrium trajectories and transition-path-sampling shooting points released by Ref.~\citenum{lazzeri2023molecular}.
After removing duplicate structures, we stratified candidates by native-contact fraction $Q$ and C$\alpha$ RMSD using bin widths of 0.05 and 0.05\,nm, respectively, and selected structurally diverse representatives by clustering interatomic-distance features within each bin. Additional configurations filled gaps on a finer grid, with greater sampling of regions where preliminary shooting results indicated intermediate committors. Reference committors were estimated using 50 unbiased shooting trials per configuration with CHARMM27 and explicit TIP3P water at 340\,K, using a 2\,fs timestep. The solvent was equilibrated with the protein fixed, and each trial began after additional solvent relaxation and a fresh draw of thermal velocities. 

We run up to 50 unbiased shooting trajectories per configuration
with CHARMM27 and explicit TIP3P water, using a 2\,fs timestep.
Basin membership is checked every 1\,ps.
Each shot initially runs for up to 50\,ns
and stops upon entering either basin.
Some shots remain unresolved.
We retain configurations for which at least 90\% of all shots reach a basin.
This leaves 2,134 test configurations.
For each configuration, the reference committor is the
fraction reaching the unfolded basin $B$ among shots
that reach either basin.
Unresolved shots are excluded from this estimate.

We use $M=32$ velocity samples per configuration.
We compare predictions from \name{} and WELP \cite{chen2026welpweightedensemble}, the SOTA semigroup method, against the same
reference shooting estimates.

\paragraph{Trp-cage.}\textit{Test configurations, shooting and baselines.}
We select 1,200 configurations from the short simulation
trajectories of Ref.~\citenum{strahan2021long}.
We divide the DGA committor range into ten bins
and select 120 configurations from each.
Training of \name{} uses independently generated trajectories.
Our training does not use any of the DGA trajectories or predictions, or the test configurations and their shooting outcomes.

From each configuration, we run 40 unbiased shooting
trajectories with fresh thermal velocities at 300\,K.
Each shot initially runs for up to 20\,ns
and stops upon entering either basin.
Shots that reach neither basin continue from their
saved positions and velocities.
The longest recorded shooting trajectory lasts
approximately 1.78\,$\mu$s, but some shots remain unresolved.
We therefore retain only configurations for which at least 36 of the 40 shots reach a basin (90\%).
This leaves 1,037 test configurations.
For each configuration, we estimate the committor as the fraction reaching the folded basin $B$ among shots that reach either basin.

We then evaluate our network on these test configurations; we use $M=16$ velocity samples per configuration.
We compare predictions from \name{}, DGA and
semigroup learning against the same reference shooting estimates.
For DGA, we follow the released implementation of
Ref.~\citenum{strahan2021long}.
Its dataset contains 1,024 trajectories of 30\,ns each,
totaling 30.72\,$\mu$s.
We follow their setting exactly and use a lag time of 12.5\,ns.

\paragraph{Ion pairs.}

\textit{Test configurations and shooting.}
We construct a separate test set for each of the six salts.
We first collect configurations from simulations
with restraints on ion separation.
For each salt, we use nine restrained separations.
After 50\,ps of equilibration at each separation,
we collect 500 configurations at intervals of 1\,ps.
We launch 16 preliminary unbiased shots from each configuration.
For starting configurations whose shots reach both basins,
we also collect configurations along those shots.
To select these path configurations, we use a preliminary
committor network trained separately with different random seeds.
This network
only guides configuration selection and hence it is not fully trained as well.
We preferentially select configurations with predicted
committors near $q=0.5$.
We combine the restrained configurations and selected
path configurations into a candidate set.
Finally, we estimate each candidate's committor using 24 unbiased shots.
We divide these estimates into 12 equal-width bins
and select 100 configurations uniformly from each bin.
This gives the final test set of 1,200 configurations per salt with roughly evenly spanning $q$.

From each selected configuration, we launch 500 new unbiased shooting trajectories with fresh thermal
velocities at 300\,K.
Each shot runs for up to 500\,ps and stops upon entering either basin. 
All final shots reach a basin.
The reference committor is the fraction reaching $B$ first.
Only these new shots contribute to the reference estimate;
the preliminary 24 shots (in the last paragraph) are used to select configurations.

We use $M=256$ velocity samples per configuration.
All of models are evaluated on the same test configurations.

\textit{Dissociation rates with RETIS.}
We calculate dissociation rates for all six networks
on all six salts using replica exchange transition
interface sampling (RETIS)\cite{vanerp2007retis}.
Interfaces are defined by $q_\theta(x)$, averaged over
16 thermal velocity samples.
We reuse the same random numbers for the velocity samples, which ensures a reproducible interface value.

The interfaces are adjusted during preliminary sampling:
we first run RETIS with initial interfaces.
After repeating these steps 300 times, we use the collected trajectories to tune the interfaces.
More precisely, for each sampled trajectory, we record the largest committor value it reaches.
This tells us whether it crosses a proposed next interface.
We use these crossing counts to place the interfaces so that approximately 1/4 of the trajectories crossing each interface reach the next one before returning to $A$.
We repeat this adjustment after another 600 and then 1,200 steps.
We then keep the interfaces fixed for the rate calculation.

\begin{figure*}[p]\centering
\includegraphics[width=0.9\textwidth]{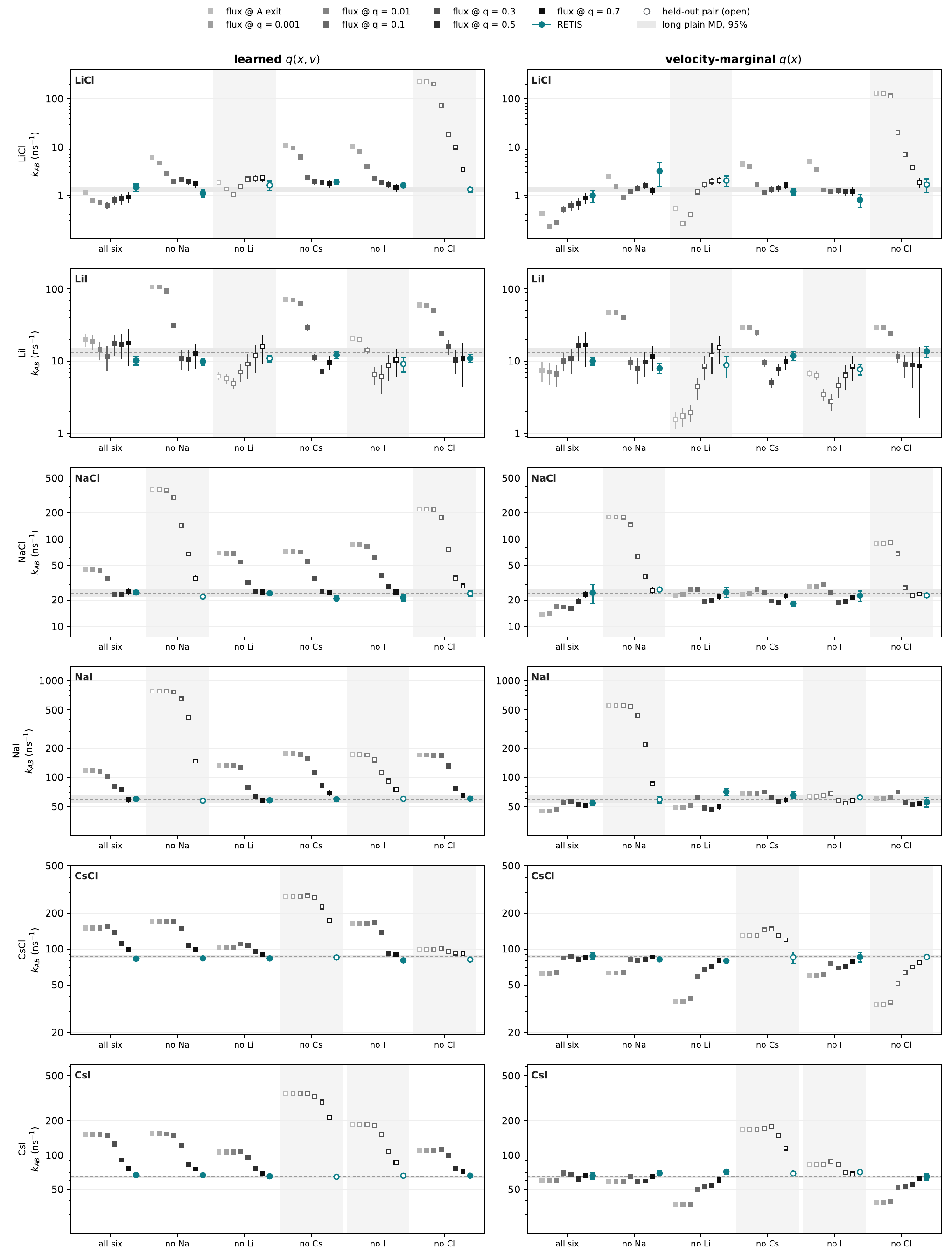}
\caption{\textbf{
{Rate} 
estimates 
{from} the 
{Hill} {relation.}} Per ion pair (rows) and network (groups), $k_{AB}$ from the flux out of $A$
weighted by the learned committor at interfaces from the exit of $A$ to $q=0.7$ (squares, light to dark), with $q(x,v)$ (left) and its
velocity average $q(x)$ (right); RETIS (teal) and long plain MD (gray band, 95\%) for reference; open symbols and shading, held-out pairs.}
\label{fig:si-ion-flux}
\end{figure*}

With the interfaces fixed, we repeat the trajectory generation and exchange steps until either ten hours of computation
have elapsed or the maximum number of repetitions is reached.
This maximum is 20,000 for the Li salts and 10,000 for the other salts.
We exclude the first 200 repetitions from the rate estimate.
Using the remaining trajectories, we esitmate the rate as
\begin{equation}
\widehat k_{AB}
=
\frac{N_{\mathrm{exit}}}{T_A}
\prod_i \widehat P_i.
\end{equation}
Here $N_{\mathrm{exit}}$ is the number of exits from $A$.
The time $T_A$ includes the time spent inside $A$ and includes time spent outside $A$ after leaving it, until the trajectory returns to $A$ or reaches $B$.
We estimate $N_{\mathrm{exit}}/T_A$ from 32 unbiased
trajectories of 10\,ns per salt.
For each interface, $\widehat P_i$ is the fraction of sampled trajectories that reach the next interface before returning to $A$.
The product includes all steps from the exit of $A$ to $B$.
We report the mean rate and its standard error from four independent runs.
For the supplementary comparison in Fig.~\ref{fig:si-ion-flux}, we repeat RETIS using interfaces defined by $q_\theta(x,v)$.
Other settings remain the same.

We obtain reference rates from longer unbiased simulations in Fig.~\ref{fig:ions}.
We count the completed transitions from $A$ to $B$ and divide this count, $N_{AB}$, by $T_A$.
The total simulated time is about 3\,$\mu$s for LiCl,
about  10\,$\mu$s for LiI and about  1\,$\mu$s for each other salt.
We calculate 95\% confidence intervals for the reference rates by treating the observed number of $A$-to-$B$ transitions
as a Poisson count.

\textit{Dissociation rates with Hill relation.}
We also estimate transition rates using the Hill relation~\cite{kramers1940brownian,baudel2023hill}.
We use the exit from $A$ and interfaces defined by
$q_\theta(x,v)=0.001$, 0.01, 0.1, 0.3, 0.5 and 0.7.

Each time a trajectory leaves $A$, we record its first
crossing of each interface before it returns to $A$
or reaches $B$.
For each interface, we sum the predicted committors
at these crossings and divide by $T_A$:
\begin{equation}
\widehat k_{AB}(\lambda)
=
\frac{1}{T_A}\sum_j q_\theta(x_j,v_j).
\label{eq:si-flux}
\end{equation}
Here $(x_j,v_j)$ gives the positions and velocities
at recorded crossing $j$.
If an interface is not reached, nothing is added to its sum.
For comparison, we repeat the calculation using $q_\theta(x)$,
averaged over 16 thermal velocity samples.

We estimate uncertainty by resampling parts of the MD
trajectories with replacement and recalculating the rate.
We repeat this 500 times and report the standard deviation
of the resulting rate estimates as the standard error.

\paragraph{Host-guest binding.}

\textit{Evaluation configurations and shooting.}

We evaluate the PaiNN model on 335 configurations from the replay buffer chosen to be approximately uniform on $h$.
Each configuration has 100 shooting trajectories with fresh
thermal velocities at 300\,K.
Of the 33,500 trajectories, 33,229 reach a basin.
For each configuration, the reference committor is the fraction
reaching $B$ among trajectories that reach either basin.
Unresolved trajectories are excluded from this estimate. During evaluation, we redraw 512 velocities per configuration from the Maxwell-Boltzmann distribution at $300K$.

\textit{Structural analysis.}

We examined how cavity hydration varied with the learned unbinding committor using release trajectories initiated from configurations in the stored SOC trajectory buffer. Starting configurations were selected by cavity-water occupancy, without using model predictions. We selected 53 distinct trajectory segments: seven segments for each target occupancy from zero to six waters and four segments for seven waters. A segment qualified for an occupancy group if at least one of its saved frames contained that number of cavity waters. The first, middle, and final configurations of each segment provided 159 starting structures. We launched four releases per structure with independently sampled, constrained Maxwell--Boltzmann velocities at 300\,K, giving 636 trajectories. Each release continued for at most 200\,ps and terminated upon reaching the bound basin, $h\leq0.65$\,nm, or the unbound basin, $h\geq1.60$\,nm. 

A water molecule was counted as "in cavity" when its oxygen lay within a cylinder of radial distance $r<0.40$\,nm from the host axis and axial coordinate $-0.10<z<0.70$\,nm relative to the cavity origin. For each configuration, we evaluated the frozen PaiNN model with explicit water inputs. To obtain a position-dependent committor, we held the coordinates fixed and averaged predictions over 512 independent draws from the constrained Maxwell--Boltzmann velocity distribution at 300\,K.

We grouped configurations by integer cavity-water count within two guest-height windows: $0.65\leq h<0.95$\,nm and $1.10\leq h<1.35$\,nm, containing 51,451 and 24,933 saved configurations, respectively. Within each window, we calculated the mean and standard deviation of the velocity-averaged committor at each water count. The plots show occupancies from zero to seven waters, retaining only bins with at least five configurations. 

\textit{Dissociation rates.}

The plotted rates use the same fixed PaiNN checkpoint.
The  RETIS estimate is 66.7\,s$^{-1}$ and the Hill
estimate is 328.8\,s$^{-1}$.

\paragraph{Hydrogen cyanide.}
\textit{Test configurations and shooting.}
We select configurations randomly from the test partition of the preparation trajectories within five intervals
of the reacting C-C distance.
The interval boundaries are
0.220, 0.235, 0.250, 0.265, 0.280 and 0.295\,nm
for liquid HCN, and
0.168, 0.173, 0.178, 0.183, 0.190 and 0.200\,nm
for the aqueous system.

Each configuration initially receives 32 unbiased shots
with fresh thermal velocities at 300\,K.
Their outcomes determine how many additional shots to run,
targeting a binomial standard error of 0.05.
Each shot stops at basin entry or after 5\,ps.
Only those configurations that have fully resolved shooting outcomes are retained for evalution.

Network predictions average over 1,024 thermal velocity draws  at 300\,K.
Both predictions and shooting estimates are expressed
as probabilities of reaching the C-C-bonded product.

\textit{Free energy profile.}
We calculate the free energy profile using umbrella sampling
along $s=\operatorname{logit}q_\theta(x)$ for the aqueous solution system.
Each prediction averages over 16 thermal velocity draws at 300\,K,
using the same random numbers at every configuration.

Window $j$ adds the potential
\begin{equation}
U_j(x)
=
\frac{Kk_{\mathrm B}T}{2}
\left[s(x)-s_j\right]^2.
\end{equation}
We use 53 window centers from $-16$ to 10,
spaced by 0.5, with $K=8$.
Each window has three independent simulations.

Each simulation runs 2\,ps of equilibration and 40\,ps of production.
After removing equilibration and the first quarter
of the production samples, we combine the windows
using the weighted histogram analysis method (WHAM).
Bootstrap resampling of the simulations within each window
gives the uncertainty.

\textit{Liquid-HCN reaction rates.}
The rate calculation describes product formation from
the free \ce{CN-} and \ce{HCNH+} ion pair.
We run plain MD of independent trajectories in parallel.
When a trajectory forms the product, neutralizes,
or forms the C-N-linked isomer, it restarts from
the state of another trajectory in the free-ion basin.

The counting uses 256 plain-MD trajectories and observes 10 product formations in 101.7\,ns.
The counting estimate is $N_{RP}/T_R$.
Here $N_{RP}$ counts transitions from the free-ion
reactant basin $R$ to the product basin $P$.
The denominator $T_R$ is the accumulated time for which
$R$ is the most recently visited basin. 
It includes the time spent outside all basins after leaving $R$, until the trajectory returns to $R$ or reaches another basin. 
A path that reaches the neutral liquid or the C-N-linked isomer counts as not reaching $P$.

The Hill relation uses the first-crossing
calculation in Eq.~\ref{eq:si-flux},
with the velocity-averaged prediction $q_\theta(x)$ at the exit from $R$ over 16 thermal velocity draws at 300\,K.
It uses a separate plain MD of 128 trajectories (3.39\,ns, 575 exits from $R$),
which also provides the exit flux and the starting paths for RETIS.
RETIS also uses interfaces defined by $q_\theta(x)$.
These predictions average over 16 thermal velocity draws
with fixed random numbers. Each RETIS run places 9 or 10 interfaces between $q_\theta(x)=10^{-4}$ and 0.6,
starting from levels at $10^{-4}$, $10^{-3}$, $3\times10^{-3}$, 0.01, 0.03, 0.1, 0.3 and 0.6
and inserting a level where fewer than 10\% of 25 trial paths reach the next one.

We perform four independent RETIS runs and average their log rates with weights equal to
the inverse squared relative standard errors,
then exponentiate the result.
The reported interval is a 95\% interval on the log-rate scale.
The counting estimate uses a 95\% Poisson interval.
Flux uncertainty uses bootstrap resampling
of trajectory and time blocks.

\textit{Illustrative transition-state configurations.}
We select examples with shooting committors near 0.5
and agreement between shooting and network predictions.
Candidates have shooting committors between 0.35 and 0.65
and absolute prediction errors below 0.2.
We prefer values between 0.4 and 0.6
with errors below 0.12.
The examples come from different preparation trajectories.

\section{Additionally analysis and experiments}
\subsection{Ion and water contributions to the committor response across salts}
\label{sec:si-ion-mech}

\begin{figure*}[!t]
\centering
\includegraphics[width=\textwidth]{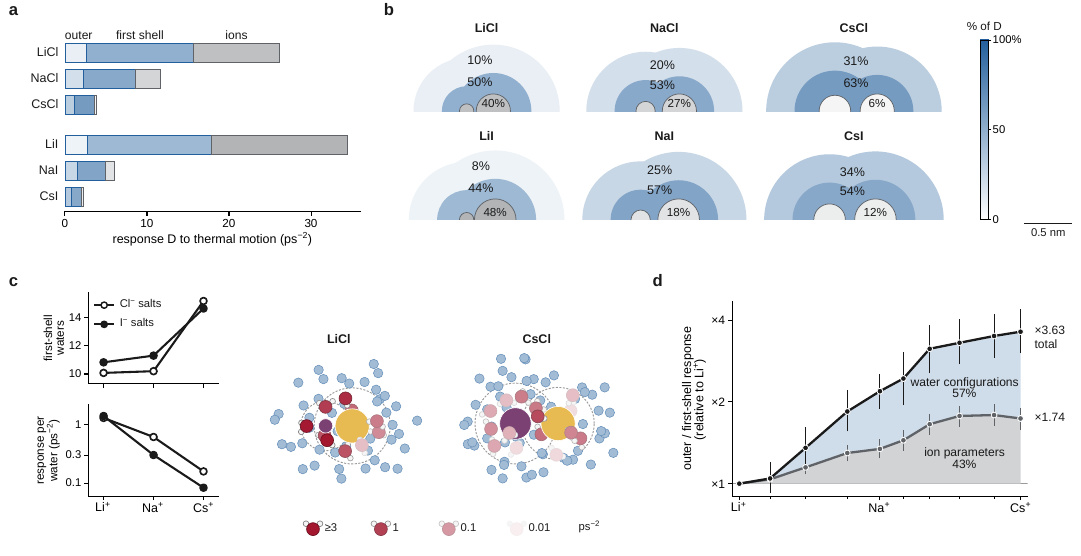}
\caption{\textbf{How ion and water motion affect the dissociation committor.}
\textbf{a},~Response of the committor to thermal motion near the transition state ($D$), separated into contributions from the ions (gray), water in the first hydration shells (blue), and water outside these shells (light blue).
\textbf{b},~The fraction of the total response contributed by each region, illustrated around the ion pairs. Darker shading indicates a larger contribution.
\textbf{c},~Number of waters in the first hydration shells (top) and their mean response per water (bottom); open and filled circles denote chloride and iodide salts. Example LiCl and CsCl configurations show first-shell waters in red, with darker colors indicating larger responses. Blue dots represent outer waters; purple and gold spheres represent the cation and anion.
\textbf{d},~Change in the relative response of outer and first-shell water from Li$^+$ to Cs$^+$. Gray shows the contribution from changing ion properties at fixed configurations, blue from changes in the sampled configurations, and black their combined effect. Error bars indicate 95\% bootstrap intervals.}
\label{fig:ion-mechanism} 
\end{figure*}

The committor learned by \name{} provides a reaction coordinate for
ion-pair dissociation, allowing us to identify which molecular motions
it is most sensitive to near the transition state.
We examine how small movements of the ions and surrounding water change
the predicted probability of dissociation near the transition state.
We use the model trained on all six salts and select test configurations
with committors between 0.4 and 0.6.
For each selected configuration, we average the network prediction
over 16 thermal velocity draws at 300\,K:
\begin{equation}
q_\theta(x)
=\frac{1}{16}\sum_{m=1}^{16}q_\theta(x,v^{(m)}),
\qquad
s(x)=\operatorname{logit}q_\theta(x).
\label{eq:si-ion-qx}
\end{equation}
The velocities are drawn from the Maxwell distribution and projected
onto the rigid water constraints.
The same draws are retained for the following analysis.

\paragraph{Contributions of ion and water motion to the dissociation committor.}
We quantify the response of the committor to  ion and water motion by the mean squared change in $q$
under thermal displacements over $\Delta t=10$\,fs.
Writing $q=q_\theta(x)$ and $s(x)=\operatorname{logit}q_\theta(x)$,
the first-order change is
\begin{equation}
\delta q\simeq q(1-q)\,\delta s,
\qquad
\delta s\simeq\nabla_x s\cdot\delta x,
\qquad
\delta x\simeq v\Delta t,
\label{eq:si-ion-logit}
\end{equation}
where $v$ denotes the thermal velocities.
We calculate the derivatives of $s$ by automatic differentiation.

For each ion $\alpha$, we use the positional gradient
$\mathbf g_\alpha=\nabla_{x_\alpha}s$.
For each water molecule $j$, we account for its rigid translation
and rotation, with derivatives $\mathbf g_j$ and
$\boldsymbol\tau_j$, respectively:
\begin{equation}
\mathbf g_j=
\sum_{a\in j}\frac{\partial s}{\partial x_a},
\qquad
\boldsymbol\tau_j=
\sum_{a\in j}(x_a-R_j)\times\frac{\partial s}{\partial x_a}
+
\sum_{m=1}^{16}\sum_{a\in j}
v_a^{(m)}\times\frac{\partial s}{\partial v_a^{(m)}}.
\label{eq:si-ion-water-gradients}
\end{equation}
Here, $R_j$ is the center of mass of water $j$.
We use the minimum-image convention to make each water molecule
whole before calculating these derivatives.
For rotation, both the atomic positions and the constrained
velocity draws used to evaluate $q_\theta(x)$ are rotated together,
giving the second term in $\boldsymbol\tau_j$.

We average the squared change in $q$ over thermal velocities.
Since the velocities of different molecules are independent and have
zero mean, the cross terms vanish.
With $w(x)=[q_\theta(x)(1-q_\theta(x))]^2$, we obtain
\begin{equation}
\left\langle(\delta q)^2\right\rangle_v
\simeq (\Delta t)^2w(x)
\left[k_{\mathrm B}T\sum_{\alpha=\pm}
\frac{|\mathbf g_\alpha|^2}{m_\alpha}
+\sum_{j=1}^{N_{\mathrm w}}H_j\right],\quad H_j=k_{\mathrm B}T
\left(\frac{|\mathbf g_j|^2}{M}
+\boldsymbol\tau_j^\top\mathbf I_j^{-1}\boldsymbol\tau_j\right).
\label{eq:si-ion-water-response}
\end{equation}
where $m_\alpha$ is the mass of ion $\alpha$.
$M$ and $\mathbf I_j$ are the mass and inertia tensor
of water $j$.
The two terms describe its rigid translation and rotation, respectively.

Averaging over the selected transition-state configurations gives
\begin{equation}
D_{\mathrm{ions}}=
k_{\mathrm B}T\left\langle
w\sum_{\alpha=\pm}\frac{|\mathbf g_\alpha|^2}{m_\alpha}
\right\rangle,
\qquad
D_R=\left\langle w\sum_{j\in R}H_j\right\rangle,
\quad R\in\{\mathrm{in},\mathrm{out}\}.
\label{eq:si-ion-contribution}
\end{equation}
Here, the brackets denote an average over configurations,
and $\mathrm{in}$ and $\mathrm{out}$ refer to water inside
and outside the first hydration shells, respectively, which we define below.
The total response is
\begin{equation}
D_{\mathrm{water}}=D_{\mathrm{in}}+D_{\mathrm{out}},
\qquad
D=D_{\mathrm{ions}}+D_{\mathrm{water}}.
\label{eq:si-ion-total-response}
\end{equation}
The response $D$ has units of ps$^{-2}$ and gives a mean squared
change in $q$ of approximately $(\Delta t)^2D$.
Supplementary Fig.~\ref{fig:ion-mechanism}a shows the three contributions, with relative contributions
$D_{\mathrm{ions}}/D$, $D_{\mathrm{in}}/D$ and $D_{\mathrm{out}}/D$.

\paragraph{Water inside and outside the first hydration shells.}
We divide water molecules into inner and outer regions using the first hydration shells of the two ions.
We define the first hydration shell using the ion-oxygen
radial distribution $g(r)$.
Its main peak indicates where nearby water molecules are
most concentrated, while the following minimum marks
the boundary of the first hydration shell.
For each salt, we calculate $g(r)$ separately for the two
ions using 1,200 test configurations, correcting the
distance histograms for spherical shell volume.

A water belongs to the inner region if its oxygen lies
within the first hydration shell of either ion, and is
counted only once.
All remaining waters belong to the outer region.
The resulting radii are listed in
Supplementary Table~\ref{tab:si-ion-water-shells}.

\begin{table}[ht]
\centering
\caption{\textbf{First hydration shell radii.}
$R_+$ and $R_-$ are the cation-oxygen and anion-oxygen
first-shell radii, respectively, in nm.}
\label{tab:si-ion-water-shells}
\begin{tabular}{lcccccc}
\toprule
Salt & LiCl & LiI & NaCl & NaI & CsCl & CsI\\
\midrule
$R_+$ & 0.2625 & 0.3125 & 0.3275 & 0.3175 & 0.4275 & 0.3925\\
$R_-$ & 0.4025 & 0.4325 & 0.3675 & 0.4175 & 0.3825 & 0.4225\\
\bottomrule
\end{tabular}
\end{table}

Supplementary Fig.~\ref{fig:ion-mechanism}a,b illustrates the
relative contributions of ions, inner water and outer water.

To distinguish the number of first-shell waters from their
individual contributions, we calculate the mean number
$\langle N_{\mathrm{in}}\rangle$ and the mean response per water,
$D_{\mathrm{in}}/\langle N_{\mathrm{in}}\rangle$.
Supplementary Fig.~\ref{fig:ion-mechanism}c compares these
quantities across the six salts.
We also show representative configurations
of LiCl and CsCl.

From Li$^+$ to Cs$^+$, the first hydration shells contain
more waters, but the response per water and the total response decrease substantially.
Outer water instead contributes a larger fraction
of the total response.
This indicates a shift in the committor's sensitivity
from ion motion and nearby water toward the surrounding solvent.

\paragraph{Does the change in relative water response come from ion properties or differences in configurations?}
The increasing contribution of outer water over inner water may arise from two effects:
Changing ion properties may alter the committor's sensitivity at a fixed configuration, or it may change the configurations found near the transition state.
To distinguish these effects, we examine
$D_{\mathrm{out}}/D_{\mathrm{in}}$ for shared configurations.

However, the transition state configurations of Li$^+$ and Cs$^+$ differ substantially.
This makes a direct comparison at fixed configurations difficult.
We therefore introduce intermediate chloride systems by
interpolating the cation mass and Lennard-Jones parameters
$(\sigma,\epsilon)$ from Li$^+$ to Na$^+$ and then to Cs$^+$,
while keeping Cl$^-$ and the water model unchanged.
This is possible because \name{} takes the ions' physical
parameters as inputs, enabling smooth intermediate chloride systems even they do not exist.
For each intermediate system, we use restrained molecular dynamics to sample configurations with network committors between 0.4 and 0.6.
We evaluate each configuration set under its own ion parameters and those of adjacent systems, using the same thermal velocity draws.
This allows us to compare the effects of ion parameters
on the same configurations while accounting for changes
in the sampled configurations between neighboring systems.
The first-shell radii are kept fixed within each part of the interpolation.

Let $X_i$ denote the configurations sampled at
$\theta_i=(m_i,\sigma_i,\epsilon_i)$.
We define
\begin{equation}
L_{ij}=
\log\frac{D_{\mathrm{out}}(X_i;\theta_j)}
{D_{\mathrm{in}}(X_i;\theta_j)},
\qquad
\Delta L_i=L_{i+1,i+1}-L_{ii}.
\label{eq:si-ion-ladder-ratio}
\end{equation}
By evaluating configuration sets under both parameter values,
we separate the change between adjacent systems into
\begin{align}
\Delta L_i^\theta
&=\frac12\left[
(L_{i,i+1}-L_{ii})
+(L_{i+1,i+1}-L_{i+1,i})
\right],\quad
\Delta L_i^X
=\frac12\left[
(L_{i+1,i}-L_{ii})
+(L_{i+1,i+1}-L_{i,i+1})
\right].
\label{eq:si-ion-ladder-split}
\end{align}
Here, $\Delta L_i^\theta$ measures the effect of changing
ion parameters while keeping configurations fixed,
whereas $\Delta L_i^X$ measures the effect of changing
configurations at fixed ion parameters.

Supplementary Fig.~\ref{fig:ion-mechanism}d shows
the cumulative changes $\exp(\sum_{i<k}\Delta L_i)$
and $\exp(\sum_{i<k}\Delta L_i^\theta)$,
relative to Li$^+$.
Error bars show 95\% intervals from 2,000 bootstrap
resamples.
Both changes in ion parameters and differences in configurations contribute to the increase in the relative response of outer water over inner water.

\providecommand{\authorcheck}[1]{%
\par\noindent\textbf{Author check.} \textit{#1}\par}

\begin{figure*}[p]\centering
\includegraphics[width=\textwidth]{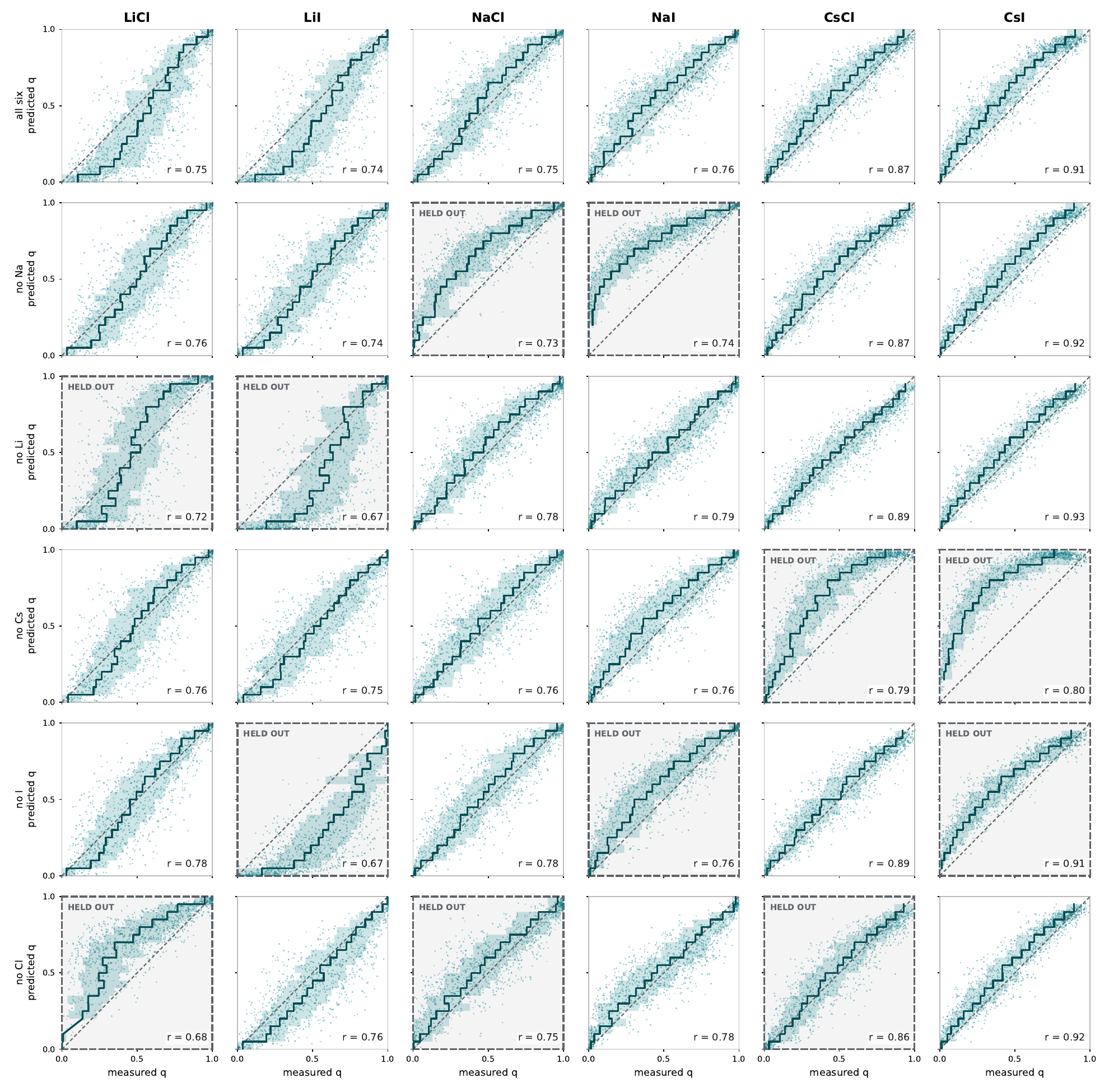}
\caption{\textbf{Committor validation of the transferable ion networks.} Rows, networks trained on all six salts or with one ion species
left out (no Na$^+$, Li$^+$, Cs$^+$, I$^-$ or Cl$^-$); columns, ion pairs. Points, the 1,200 test configurations per pair: shooting
committor (500 shots with Maxwell velocities) against $q(x)$ (256 velocity draws). Line, mean shooting committor in 0.05-wide bins of the
prediction; shading, $\pm$1 s.d.; dashed, identity; dashed frames, held-out pairs.}
\label{fig:si-ion-qx}
\end{figure*}

\bibliography{references}

\end{document}